\documentclass[fleqn,usenatbib]{mnras}

\usepackage{newtxtext,newtxmath}
\usepackage[T1]{fontenc}
\usepackage{ae,aecompl,graphicx, amsmath, natbib} % amssymb, 

\title[Contrastive learning of TNG-Cluster in X-ray]{ERGO-ML: A continuous organization of the X-ray galaxy cluster population in TNG-Cluster with contrastive learning}

\author[U. Chadayammuri et al.]{Urmila Chadayammuri$^{1}$\thanks{E-mail:chadayammuri@mpia.de},
    Lukas Eisert$^{1}$,
    Annalisa Pillepich$^{1}$,
    Katrin Lehle$^{2}$, \newauthor
    Mohammadreza Ayromlou$^{2,3}$ and  
    Dylan Nelson$^{2}$
\\\\
$^{1}$ Max Planck Institut f\"ur Astronomie, K\"onigstuhl 17, 69121 Heidelberg, Germany \\
$^{2}$ Universit\"{a}t Heidelberg, Zentrum f\"{u}r Astronomie, ITA, Albert-Ueberle-Str. 2, 69120 Heidelberg, Germany \\
$^{3}$ Argelander-Institut f\"ur Astronomie, Auf dem H\"ugel 71, D-53121 Bonn, Germany\\
}

\pubyear{2024}

\begin{document}
\label{firstpage}
\pagerange{\pageref{firstpage}--\pageref{lastpage}}
\maketitle

\begin{abstract}
The physical properties of the intracluster medium (ICM) reflect signatures of the underlying gravitational potential, mergers and strong interactions with other halos and satellite galaxies, as well as galactic feedback from supernovae and supermassive black holes (SMBHs). Traditionally, clusters have been characterized in terms of summary statistics, such as halo mass, X-ray luminosity, cool-core state, luminosity of active galactic nuclei (AGN), and number of merging components. In this paper of the Extracting Reality from Galaxy Observables with Machine Learning series (ERGO-ML), we instead consider the full information content available in maps of X-ray emission from the ICM. We employ Nearest Neighbour Contrastive Learning (NNCLR) to identify and populate a low-dimensional representation space of such images. Using idealized X-ray maps of the 352 clusters of the TNG-Cluster cosmological magnetohydrodynamical simulation suite, we take three orthogonal projections of each cluster at eight snapshots within the redshift range $0\leq z<1$, resulting in a dataset of $\sim$8,000 images. Our findings reveal that this representation space forms a continuous distribution from relaxed to merging objects, and from centrally-peaked to flat emission profiles. The representation also exhibits clear trends with redshift, with halo, gas, stellar, and SMBH mass, with time since a last major merger, and with indicators of dynamical state. We show that an 8-dimensional representation can be used to predict a variety of cluster properties, find analogs, and identify correlations between physical properties, thereby suggesting causal relationships. Our analysis demonstrates that contrastive learning is a powerful tool for characterizing galaxy clusters from their images alone, allowing us to derive constraints on their physical properties and formation histories using cosmological hydrodynamical galaxy simulations.
\end{abstract}


%%%%%%%%%%%%%%%%%%%%%%%%%%%%%%%%%%%%%%%%%%%%%%%%%%%%%%%%%%%%%%%%

\section{Introduction}

The make-up of galaxy clusters is determined by cosmological structure evolution, on the one hand, and feedback from constituent galaxies on the other. Perhaps nowhere do these two factors leave clearer signatures than in the intracluster medium (ICM), the diffuse, X-ray luminous plasma that permeates the space among cluster galaxies. Decades of X-ray observations with high angular-resolution instruments like \textit{Chandra} and \textit{XMM-Newton} have revealed a diversity of thermodynamic structures, from cool- to non-cool cores \citep{Peres1998, Chen2007, McDonald2013}, from relatively isolated to strongly merging systems \citep{Markevitch1999}, from smooth gas distribution to ICM permeated with cold and shock fronts \citep{Ascasibar2006}, and with various levels of activity from active galactic nuclei \citep[AGN, ][]{Nulsen2005,Fabian2006, Birzan2008, Panagoulia2014}. The ongoing eROSITA All-Sky Survey has increased the X-ray cluster sample to an unprecedented 12,247 \citep{Bulbul2024}, albeit at lower spatial resolution than with \textit{Chandra} and XMM-Newton. The recently launched XRISM is complementing these imaging studies with line-of-sight (LOS) velocity measurements for the gas enabled by high spectral resolution \citep{Zhuravleva2024}. Proposed upcoming surveys like New-Athena \citep{Barret2023}, Line Emission Mapper (LEM, \citet{Kraft2022}), and Lynx \citep{Schwartz2019} will provide sharper images with even finer spectral resolution.  

For all but the most relaxed galaxy clusters observed at high spatial resolution by e.g. \textit{Chandra} and \textit{XMM-Newton}, it is extremely challenging to infer cluster properties such as the mass and baryon fraction from the ICM emission. At lower resolution, it is challenging to tell if a galaxy cluster has a cool or non-cool core, shock or cold fronts from recent mergers, or AGN-driven cavities.

The ICM is intrinsically complex. The emission itself comes from multiple cooling channels - thermal bremsstrahlung, inverse Compton scattering, and line emission from various elements --, each leaving distinct signatures that must be carefully distinguished to accurately infer the underlying physical conditions. Furthermore, non-thermal processes, such as turbulent and bulk motions and cosmic-ray pressure, can be difficult to separate from thermal broadening when spatial and/or spectral resolution are limited \citep{Rebusco2008, Ogorzalek2017, Lau2017, Cucchetti2019}. 

Beyond instrumental considerations, there is the challenge of capturing the complex information in a 2D image. Models of the ICM thermodynamic structure, gas dynamics, and magnetic field necessarily rely on a variety of assumptions and simplifications, introducing uncertainties that propagate into the interpretation of X-ray data. Eventually, these models reduce complex images to radial profiles \citep{Vikhlinin2005, Piffaretti2005, Voit2005} or scalar summary statistics \citep{Vikhlinin2006, Chen2007,Maughan2008, Ghirardini2022}. Scaling relations between these scalars are helpful, but carry a large amount of scatter \citep{Pratt2009}. These reduced metrics are then used to infer relationships among physical processes, such as AGN feedback, cluster mergers, cooling flows, and star formation \citep{Rafferty2008, Sanderson2009,Maughan2012, McDonald2013, McDonald2014}, or even cosmological parameters \citep[e.g.][]{Ghirardini2022}. Indeed, these simplified models have been very useful in building our understanding of the evolution of galaxy clusters and of the whole Universe. 

Cosmological hydrodynamical simulations of galaxy clusters play a crucial role in understanding the complex processes that govern cluster formation and evolution. By these we mean simulations that incorporate, in addition to gravity and gas dynamics, also various astrophysical processes, --  such as gas cooling, star formation, and feedback from AGN -- in order to model the ICM and its interaction with dark matter. Mock observations of simulated clusters can, on the one hand, validate the realism of the physical models assumed in the simulations and, on the other, help interpret complex observations of the ICM. Notably, such comparisons have helped quantifying the hydrostatic-mass bias \citep{Rasia2012, Ota2018, Barnes2021, Braspenning2024}, estimating ICM turbulence \citep[][and references therein]{Zhuravleva2023, Truong2024}, inferring velocity structures from X-ray observations \citep{Biffi2013}, and the effect of AGN contamination on the detectability of galaxy clusters in surveys \citep{Biffi2018}. In addition, they provide the initial conditions and statistical context for more idealized simulation setups, which are key to interpreting individual systems like Perseus \citep{Lau2017, Truong2024}, the Bullet Cluster \citep{Markevitch2002, Markevitch2004, Clowe2004}, Abell 2146 \citep{Chadayammuri2022} and others \citep{ZuHone2018}.

In parallel, machine learning techniques offer a complementary set of tools to interpret ICM images, bypassing the need for summary statistics. Convolutional Neural Networks (CNNs) have been very powerful in inferring galaxy cluster mass from observations \citep{Ho2019, Ntampaka2019, Kodi2021, Krippendorf2024}; the introduction of variational and Bayesian CNNs now allows to quantify the uncertainties in these predictions \citep{Ho2021}. Furthermore, they can classify images based on dynamical state \citep{Arendt2024}. In the inverse problem, CNNs can also be used to predict observable i.e. baryonic images of galaxy clusters given dark matter-only inputs from N-body simulations \citep[e.g.][]{Thiele2020, Chadayammuri2023, Anbajagane2024}, offering an alternative, and once that is spatially resolved, to baryon post-processing tools based on semi-analytic and abundance-matching models. Crucially, CNNs offer the additional benefit that they naturally reproduce the scatter in the scaling relations -- this tells us that the scatter is in fact a result of the diverse spatial structure within galaxy clusters and of their evolutionary history.

A key limitation of CNNs is that they are not invariant to transformations that are common in astronomical (or, for that matter, any) observations - translation, rotation, dynamical-range a.k.a.surface-brightness limits, zooming and cropping. However, it is possible to implement such invariances by applying appropriate operations as augmentations to a training set and then training the CNN on the augmented sample \citep{Dieleman2015}. Rotational invariance in particular has been implemented for a wide variety of classification problems in astronomy \cite{Cabrera-Vives2017, Schaefer2018, Zhu2019, Hosenie2021,Scaife2021}.  

Contrastive learning takes the principle of augmentation one step further. This architecture aims to learn possible representations by maximizing the similarity between positive pairs (e.g., different augmentations of the same image) while minimizing the similarity between negative pairs (e.g., images of different objects) \citep{Chen2020}. Nearest Neighbour Contrastive Learning \citep[NNCLR, ][]{Dwibedi2021} further enhances the self-supervised organisation of the images by selecting a random subset of the sample as a ``support set''. Two augmentations ($a_1,a_2$) of an image ($a_0$) are created, and the then the nearest neighbour $n_1$ of $a_1$ is found in the support set. The cost function to be minimised is then the distance in the representation space between $n_1$ and $a_2$ (rather than between $a_1$ and $a_2$ in conventional contrastive learning). Not only is the ensuing representation invariant to the implemented augmentations, but it sorts the entire population of the training sample by similarity. 

In the context of the ERGO-ML project (Extracting Reality from Galaxy Observables with Machine Learning), \citet{Eisert2024} trained a NNCLR architecture using mock photometric images of galaxies from the IllustrisTNG simulations and showed that the resulting representation encodes features, not only of the galaxies' morphology, but also of their stellar assembly and merger history. The ICM in clusters, however, poses a different set of advantages and challenges than optical/infrared observations at the galaxy scale \citep[c.f.][]{Cavaliere1976}. On the one hand, the ICM is optically thin, so that obscuration effects are negligible. It is also more spherically symmetric, so that the dependence on viewing angle is reduced (although far from eliminated, due to halo triaxiality, mergers, feedback, etc). On the other hand, observations of the ICM are primarily at X-ray wavelengths, where photons are scarce, and spectral and/or spatial resolution are limited \citep{Cavaliere1976, Arnaud2005}. Instruments like Chandra and XMM-Newton have observed of order a hundred galaxy clusters, which is too small a sample to train neural networks with. X-ray surveys of galaxy clusters with ROSAT \citep{Ebeling1998, Bohringer2004} and eROSITA \citep{Liu2022, Bulbul2024} offer significantly larger samples, albeit at lower exposure time and spatial resolution, leading to the loss of potentially-informative features for the bulk of the sample. 

The physical questions of interest are also slightly different between galactic and galaxy-cluster scales. What is the origin of the variety of cores in galaxy clusters, ranging from cool- to non-cool, and is there a better way to describe such diversity as a bimodality? What is the recent and cumulative history of AGN activity in galaxy clusters? Does star formation, and its history, leave a mark? Did any given cluster recently undergo a merger, and if so, when, at what mass ratio, and with how much angular momentum? The training needs to be different from that of a galaxy image catalog to account for the different questions of interest. 

\begin{figure*}
    \includegraphics[trim=0cm 0.52cm 0cm 0cm, clip, width=\textwidth]{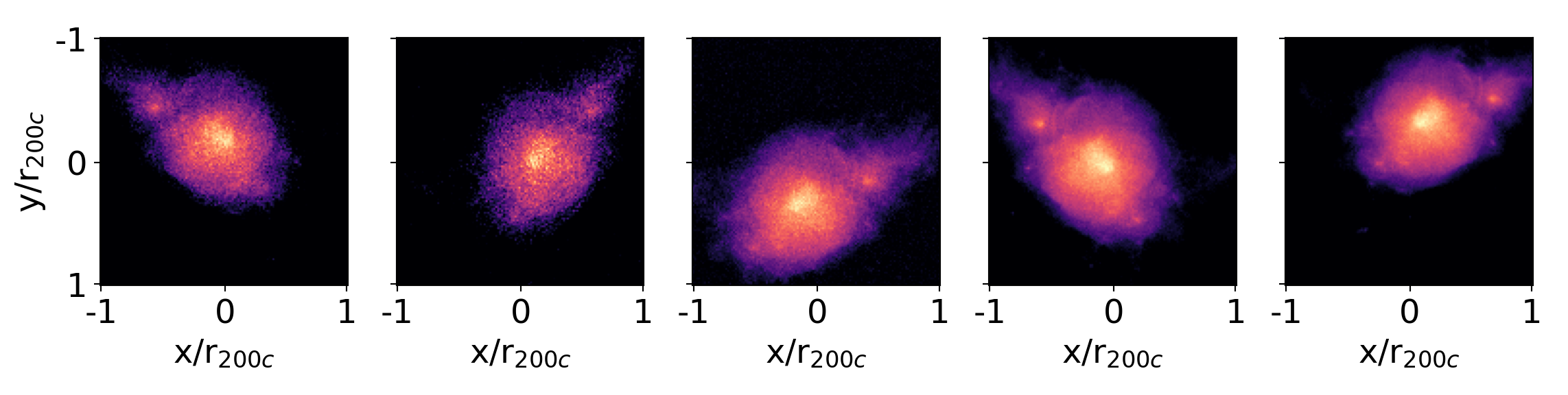}
    \caption{\textbf{The foundations of the contrastive learning approach used throughout.} We show the five augmentations implemented in our contrastive learning method for a randomly-chosen TNG-Cluster object, i.e. a simulated galaxy cluster. Each augmentation is a combination of the transformations described in \S \ref{sec:nnclr}; the magnitude of zoom, rotation, translation, blurring, noise, and surface brightness clipping are all drawn randomly within the ranges specified in the text. Similarly, there is a 50 percent probability that a horizontal or vertical flip is implemented. In practice, only two augmentations of a given raw image are produced for a single training epoch.}
    \label{fig:augment}
\end{figure*}

Here, we build upon the contrastive learning architecture developed in \citet{Eisert2024} and optimise it for X-ray observations of the ICM. We hence apply it to thousands of idealized X-ray maps of simulated galaxy clusters across a wide range of total mass, redshift, and evolutionary stages. These are taken from the novel suite of cosmological magnetohydrodynamical (MHD) simulation of galaxies, and massive galaxy clusters, known as TNG-Cluster \citep{Nelson2024}. The aim is to see whether it is possible to sort and organize X-ray maps of galaxy clusters so that they encode information about the underlying physical state and past assembly history. Crucially, we implement proxies for exposure time and spatial resolution as augmentations in the training process to evaluate the possible effects of different observational strategies, in addition to e.g. rotation, translation, Gaussian noise, and flipping that were already implemented in \citet{Eisert2024}. 

\S \ref{sec:tng} describes the TNG-Cluster simulation, and how the training images were extracted from this simulated sample. \S \ref{sec:nnclr} describes the NNCLR algorithm, and our customisations to it. \S \ref{sec:results} presents the self-organised atlas of galaxy cluster images, and explores how galaxy cluster properties are distributed in this space. It also highlights several applications of contrastive learning in the study of galaxy clusters. \S \ref{sec:discuss} discusses limitations and future work. We close with conclusions in \S \ref{sec:concl}.

%%%%%%%%%%%%%%%%%%%%%%%%%%%%%%%%%%%%%%%%%%%%%%%%%%%%%%%%%%%%%%%%

\section{Methods}

The paper is based upon the output of the TNG-Cluster simulation, intrinsic maps of the X-ray emission generated from it, contrastive learning, and dimensionality reduction for both visualization and analysis purposes. In the following, we hence describe in detail these components of the work. 

\subsection{The TNG-Cluster simulation suite}
\label{sec:tng}

TNG-Cluster\footnote{\url{www.tng-project.org/cluster/}} \citep{Nelson2024} is a suite of zoom-in simulations recently added to the IllustrisTNG family of cosmological simulations \citep{Springel2018, Pillepich2018b, Pillepich2019, Marinacci2018, Naiman2018, Nelson2018, Nelson2019a, Nelson2019}. Its main objective is to significantly increase the statistical sampling of the most massive and rare gravitationally-bound structures in the Universe: galaxy clusters, with a special focus to those with halo masses greater than $10^{15}M_\odot$. The TNG-Cluster project re-simulates 352 cluster regions drawn from a 1 Gpc$^3$ volume, which is thirty-six times larger than the original TNG300 simulation. 

Overall, the TNG simulations\footnote{Data from the TNG simulations, including value-added catalogues, are publicly available at \url{www.tng-project.org/data/} \citep{Nelson2019}.} aim to provide insights into the formation and evolution of galaxies and large-scale structures in the Universe. They use the moving mesh code AREPO \citep{Springel2010} to solve the equations of gravity and ideal MHD within an expanding LCDM Universe. The TNG galaxy formation model \citep{Weinberger2017, Pillepich2018} incorporates many important astrophysical processes relevant for galaxy formation, including star formation, and the evolution of and feedback from stars. It tracks chemical enrichment from multiple channels, including Type Ia and Type II supernovae (SNe), winds from massive stars, and NS-NS mergers. TNG also models the seeding, accretion, merging, and feedback from supermassive black holes (SMBHs). All runs include magnetohydrodynamics by following the self-consistent amplification of a vanishingly-small primordial seed field \citep{Pakmor2011}.

\begin{figure*}
    \includegraphics[width=\textwidth]{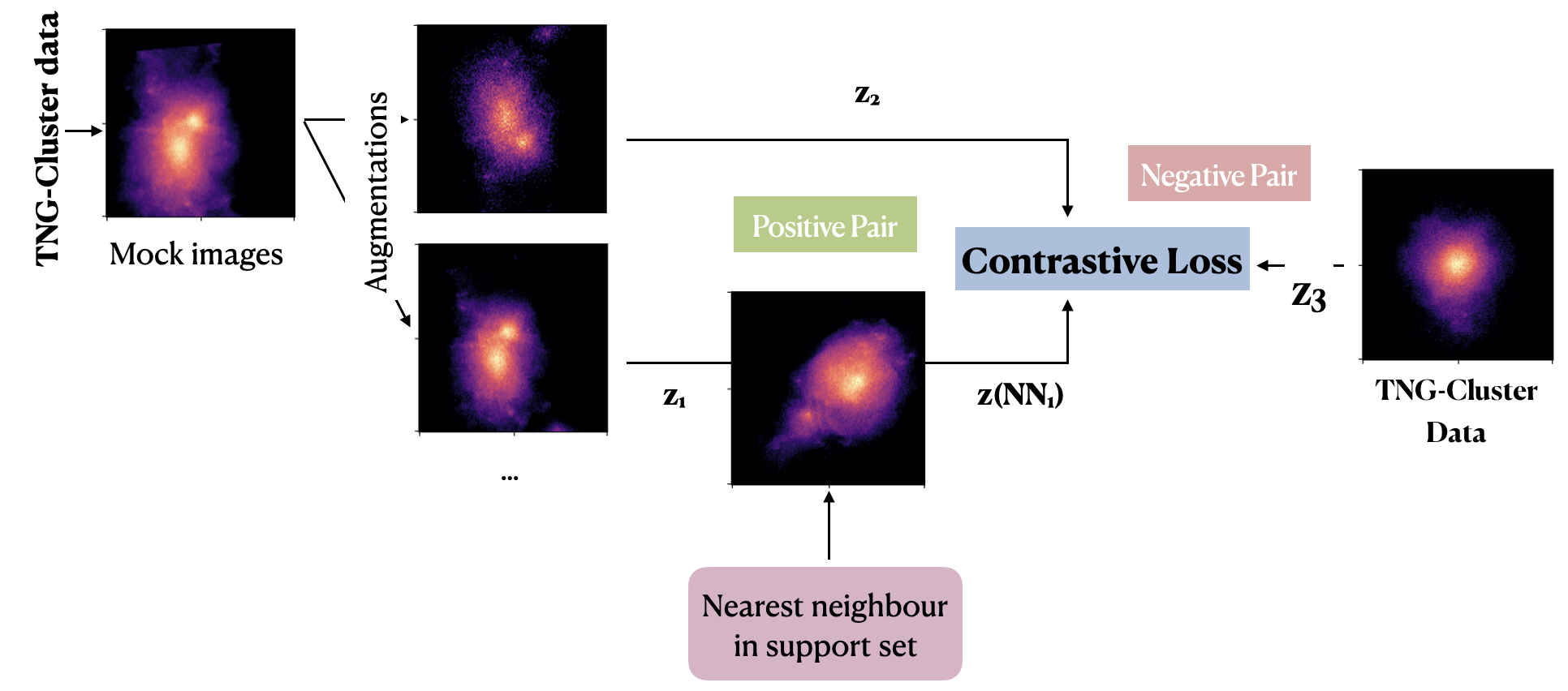}
    \caption{\textbf{Schematic of the NNCLR training architecture adopted throughout.} First, we take the projected X-ray emission map of a halo in TNG-Cluster, at $z \leq 1$ and $M_{\rm 200c} \ge 10^{14.3}M_\odot$. Then, we apply two augmentations to it, each a combination of those described in \S \ref{sec:nnclr}. We compute the latent representation of these two augmentations (z$_1$ and z$_2$), given by the weights and biases of the neural network at this training epoch. For one of the augmentations, we find the image in the support set with the most similar representation, z(NN$_1$). Meanwhile, some other training image, also with a random set of augmentations applied to it, is chosen as the negative counterpart; using the same weights and biases, its representation is z$_3$. The contrastive loss function, given in Eq.~\ref{eq:infonce}, minimises the distance between z$_2$ and z(NN$_1)$ while maximising the distance between z$_2$ and z$_3$.}
    \label{fig:nnclr}
\end{figure*}

\subsection{Modeling X-ray emission and the cluster sample}
\label{sec:emission-model}

X-ray emission in galaxy clusters is dominated by bremsstrahlung (aka free-free) emission, which has negligible metallicity-dependence. However, there is also a contribution from high-ionisation metal lines such as O, Mg, Si and Fe \citep{Matsushita2007}. Intrinsic X-ray emission maps for the TNG-Cluster sample have been made \citep{Nelson2024} in several different configurations. The maps have a field of view of $4R_{200c}$ and a line-of-sight depth of $2R_{200c}$. Using the density, temperature and metallicity of each gas cell, the intrinsic X-ray emissivity is derived using APEC \citep{Smith2001}, and projected through the adaptively-sized gas cells using the usual cubic-spline approach \citep[see][for details]{Nelson2024}.

Projections are taken along three orthogonal lines of sight, namely the $\hat{x}, \hat{y}, \hat{z}$ axes of the simulation box. Since the clusters are randomly oriented within the box, this means that the images have random viewing angles. Due to their significant triaxiality, the three projections are different enough to act as independent samples. Fig.~\ref{fig:example-projection} shows examples of these X-ray maps for a single halo across three different redshifts.

From the TNG-Cluster output, we consider the 352 primary-zoom halos, which have $M_{\rm 200c} = 10^{14.3 - 15.4} M_\odot$ at $z=0$, and their progenitors at eight snapshots within the redshift range $0\leq z<1$. The final image sample consists of 7968 maps.

\subsection{The NNCLR Architecture}
\label{sec:nnclr}

A review of the Nearest Neighbour Contrastive Learning (NNCLR) architecture used in this work can be found in \citet{Eisert2024}; below, we cover only the details pertinent to the presented analyses. 

We start with the images of the logarithm of the intrinsic X-ray emission from TNG-Cluster halos at $0 \le z \le 1$. Each map, described in \S \ref{sec:emission-model}, is normalised such that the value of the 99-th percentile pixel in the central 10 percent of the image is set to 1, and anything 4 orders of magnitude or more fainter than that is set to 0. Note that this means that the algorithm does not know whether one image is brighter or fainter than another overall. 

The sample is split, as per usual, into an 80/10/10 training/validation/test sets. Images in the training set are passed through a set of data augmentation transformations, creating multiple augmented views of each original sample. These augmented views serve as different ``views'' of the same underlying data, capturing different aspects of its structure. We implement:
\begin{itemize}
    \item Zoom: by up to 10 percent;
    \item Rotation: by up to $\pm 90^\circ$;
    \item Affine translation: by up to 25 percent in either direction;
    \item Horizontal/vertical flip: implemented randomly;
    \item Blur: convolution with a Gaussian kernel with width in pixels drawn randomly from the $[10^{-3}, 1]$ range;
    \item Noise: addition of Gaussian noise with mean and width drawn randomly from the $[0.01, 0.08]$ ranges;
    \item Clipping (\textit{optional}): in a subset of the models, we additionally vary the value of the faintest visible pixel. This choice models the fact that X-ray observations have widely varying exposures, producing images with dynamic ranges. 
\end{itemize}

As an example, five random augmentations of the TNG-Cluster halo with ID 19198592 at $z = 0.90$ are shown in Fig.~\ref{fig:augment}. 

The augmented views of the data are then passed through a Residual Network (ResNet), which acts as a feature extractor. This is a modification of a CNN that aims to predict the difference, i.e. residual, between layers, rather than the layer itself, in order to tackle the vanishing gradient problem that otherwise plagues deep neural networks \citep{Wu2019}. The ResNet maps each view of the data to a sparse representation, known as the embedding space. We experiment with the size of the representation, using N = 2, 4, 8, 16, 64, 128 and 256, and choose as default a dimension of 8. In the embedding space, NNCLR applies a contrastive objective function to encourage similar views to be closer to each other while pushing dissimilar views further apart. Specifically, it uses a nearest neighbor contrastive loss, where for each view, the loss is computed based on the distance between the embeddings of that view and its nearest neighbor among the embeddings of other views of the same sample \citep{Dwibedi2021}:

\begin{equation}
\label{eq:infonce}
\mathcal{L}_{i} = -  \log{
	{
		\frac{\exp{(z_i\cdot z_i^+/\tau)}}
		{\exp{(z_i \cdot z_i^+ / \tau)}
			+ \sum_{z^- \in \mathcal{N}_i}{\exp{(z_i \cdot z^-/\tau )}}}
	}
} 
\end{equation}

where ($z_i$, $z_i^+$) is the positive pair, ($z_i$, $z^-$) is a negative pair and $\tau = 0.04$ regulates the soft-max activation. A visual schematic description of the NNCLR approach adopted in this work is given in Fig.~\ref{fig:nnclr}.

\begin{table*}
    \begin{tabular}{l|p{5.1in}|c}
    \hline
        Name & Description for any given cluster & Units \\
        \hline
        z & Redshift of the simulation snapshot & -\\
        $M_{\rm 200c}$ & Total cluster mass, i.e. all the mass within a sphere whose mean density is 200 times the critical density of the Universe, at the time the halo is considered.  & $\log(M_{\odot})$ \\
        $M_*$ & Total BCG stellar mass $^\dagger$ & $\log(M_\odot)$ \\
        $M_{\rm gas}$ & Total halo gas mass $^\dagger$ & $\log(M_\odot)$ \\
        $M_\bullet$ & Mass of the central SMBH $^\dagger$ & $\log(\rm{M}_\odot)$ \\
        BHAR & Instantaneous accretion rate of the central SMBH $^\dagger$ & $\log(\rm{M}_\odot\,\rm{Gyr}^{-1})$ \\
        $E_{\rm therm, cum}$ & Total energy injected by the central SMBH in the thermal mode since formation $^\dagger$ & erg \\
        $E_{\rm kin, cum}$ & Total energy injected by the central SMBH in the kinetic mode since formation $^\dagger$ & erg \\
        SFR & SFR within the half-mass radius of the BCG & $\log(\rm{M}_\odot\,\rm{yr}^{-1})$ \\
        $t_{\rm merger}$ & Time since the last halo-halo merger with a mass ratio R $\geq 0.2$ & Gyr \\
        $t_{\rm BCGmerger}$ & Time since the BCG last merged with another galaxy with a mass ratio R $\geq 0.2$ & Gyr \\
        COM Offset & Offset between the center of mass of the halo and the position of its most bound particle \citep{Ayromlou2023} & $R_{\rm 200c}$ \\
        $C_{\rm phys}$ & X-ray luminosity concentration $L_X(< 40 $kpc$)/L_X(<400 $kpc$)$ \citep{Lehle2024} & - \\
        $t_{\rm cool, c}$ & Central cooling time -- measured within $0.012R_{\rm 500c} \sim 0.008R_{\rm 200c}$ \citep{Lehle2024} & Gyr \\
        $n_{e} (r=0)$ & Central electron number ensity -- measured within $0.012R_{\rm 500c} \sim 0.008R_{\rm 200c}$ \citep{Lehle2024} & cm $^{-3}$\\
        $\alpha_e(r=R_{\rm 500c})$ & Slope of the electron number density at $0.04R_{\rm 500c} \sim 0.026R_{\rm 200c}$ \\
    \hline
    \end{tabular}
    \caption{\textbf{Properties of the galaxy clusters considered in this work}. The properties denoted with a $^\dagger$ symbol are shown in Fig.~\ref{fig:umap-mnorm} and considered from there onwards as offsets from the mean value for halos with $M_{\rm 200c}$ within 0.1 dex of the corresponding halo. The central SMBH is the most massive SMBH of the central galaxy of the cluster i.e. of the Brightest Cluster Galaxy (BCG). Halos are defined as linked structures identified using the Friends-of-Friends (FoF) algorithm. The central cooling time, the electron number density slope, and the X-ray luminosity concentrations are, respectively, smaller, larger, and larger for CCs than for NCCs. The CoM-Peak offset and the mass fraction in the BCG are, respectively, smaller and larger for relaxed than for unrelaxed clusters. }
    \label{tab:props}
\end{table*}

\subsection{UMAP}

The representation of the images produced by NNCLR is compact (we ran models with between 2 and 256 dimensions) compared to the input image size (256 $\times$ 256 = 65536 pixels per map), but it is still too high dimensional for visual inspection and interpretation. UMAP (Uniform Manifold Approximation and Projection, \citet{McInnes2018}) is a method for reducing the dimensionality of a manifold, by using robust and proved theorems in topology. We will make extensive usage of UMAPs for both inspection and validation as well as for scientific inference.

UMAP is implemented through the python package \texttt{umap}\footnote{\url{https://umap-learn.readthedocs.io}}. The geometry of the UMAP depends on the number of neighbours used to compute the local geometry of the manifold (here, the representation space from the NNCLR) and the minimum distance between the reduced mappings of similar points. UMAP with low values of \texttt{n\_neighbour} will convey more about the local rather than the global structure of the manifold, whereas lower \texttt{min\_dist} yields a clumpier UMAP. 

Since our goal is clear visualisation, we reduce the representation to a 2D UMAP. We test values of $10 < n\_\texttt{neighbour} < 200$ and $0.1 < \texttt{min\_dist} < 1$ and pick a combination by eye that produces a smooth UMAP that fills up the available coordinate space. For the fiducial model, we adopt $n\_\texttt{neighbour} = 50$ and $\texttt{min\_dist} = 0.25$. 

\subsection{Properties of galaxy clusters}

The self-supervised sorting adopted in this paper only uses images as input. However, after the fact, we do explore how various properties of the simulated clusters are distributed in the representation space, and how well they can be described by it. We summarise the cluster properties we focus on in Table \ref{tab:props}. 

\begin{figure*}
    \centering
    \includegraphics[width=1\textwidth]{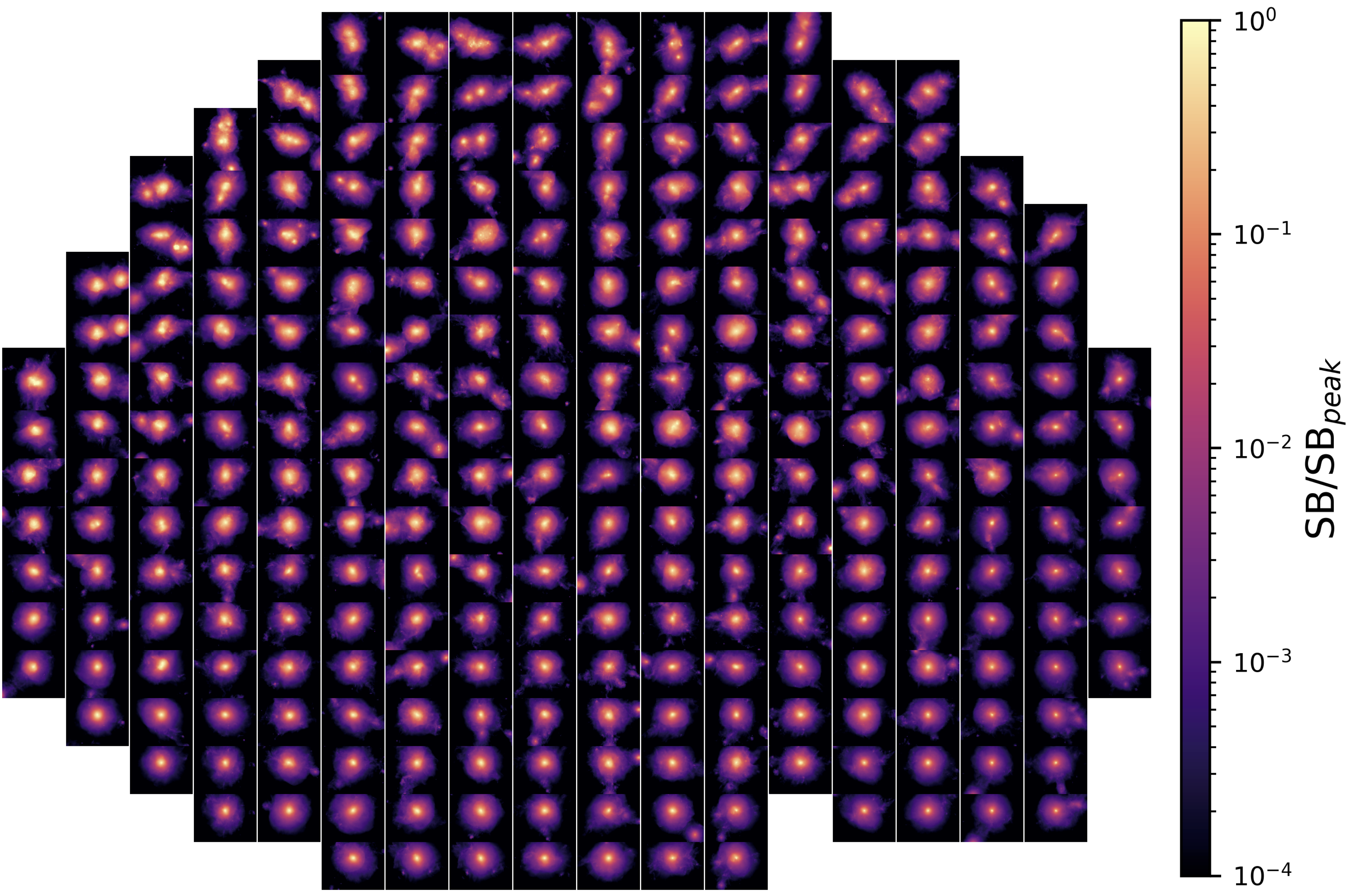}
    \caption{\textbf{Outcome of the NNCLR method applied to TNG-Cluster simulated systems, visualized in 2 dimensions.} We show a subset of the images from the TNG-Cluster sample analyzed by our NNCLR method and sorted by their UMAP coordinates. Clear trends can be seen in this representation of the X-ray maps, indicating that the contrastive learning technique organizes the cluster maps in meaningful ways: merging clusters with well-separated components are at the top, moving progressively towards smooth, isolated objects at the bottom. Furthermore, objects to the bottom right appear more centrally peaked, while those to the bottom left have flatter profiles. Note that the images are all self-normalised, i.e. the brightest pixel in each image has a value of 1 and the faintest a value of 0. Therefore, information about total or relative luminosity is removed before training. }
    \label{fig:sorted-images}
\end{figure*}

Total (gas) cluster mass is computed by summing up all resolution elements (all cells) belonging to the Friends-of-Friends halo of any given cluster. By stellar mass, we denote the gravitationally-bound stellar mass of the brightest cluster galaxy (BCG), the most massive and central {\it subhalo} of the cluster. The black-hole properties all refer to the most massive SMBH in the BCG. Stellar, gas, and SMBH masses are all expected to scale with the halo mass; therefore, after a first time, we show these properties normalised by the halo mass. Since the SMBH accretion and feedback (a.k.a. energy injection) rates are further tied to the SMBH mass, we show these in comparison to the mean value for halos within 0.1 dex in $\log(M_{\rm 200c})$. The star formation rate (SFR) is the instantaneous one of the BCG. 

Measures of the thermodynamic state of the cluster core were computed by \citet{Lehle2024} and indicators of the cluster dynamical state by \citet{Ayromlou2023}. The core properties we leave in physical units, because there is a wide variety at all halo masses. 

Merger trees are calculated using LHaloTree \citep{Springel2005}, which connects halos between consecutive snapshots and hence across time. For each cluster, we  also identify all merger events of the central BCG. For each merger, we compute the mass ratios of the total mass within $R_{\rm 200c}$. We focus in particular on the time since the last major merger, whereby we require a mass ratio $ > 0.2$ and consider separately halo-halo and BCG-galaxy mergers.

%%%%%%%%%%%%%%%%%%%%%%%%%%%%%%%%%%%%%%%%%%%%%%%%%%%%%%%%%%%%%%%%

\begin{figure*}
    \centering
    \includegraphics[trim=0cm 1cm 0cm 0cm, clip, width=\textwidth]{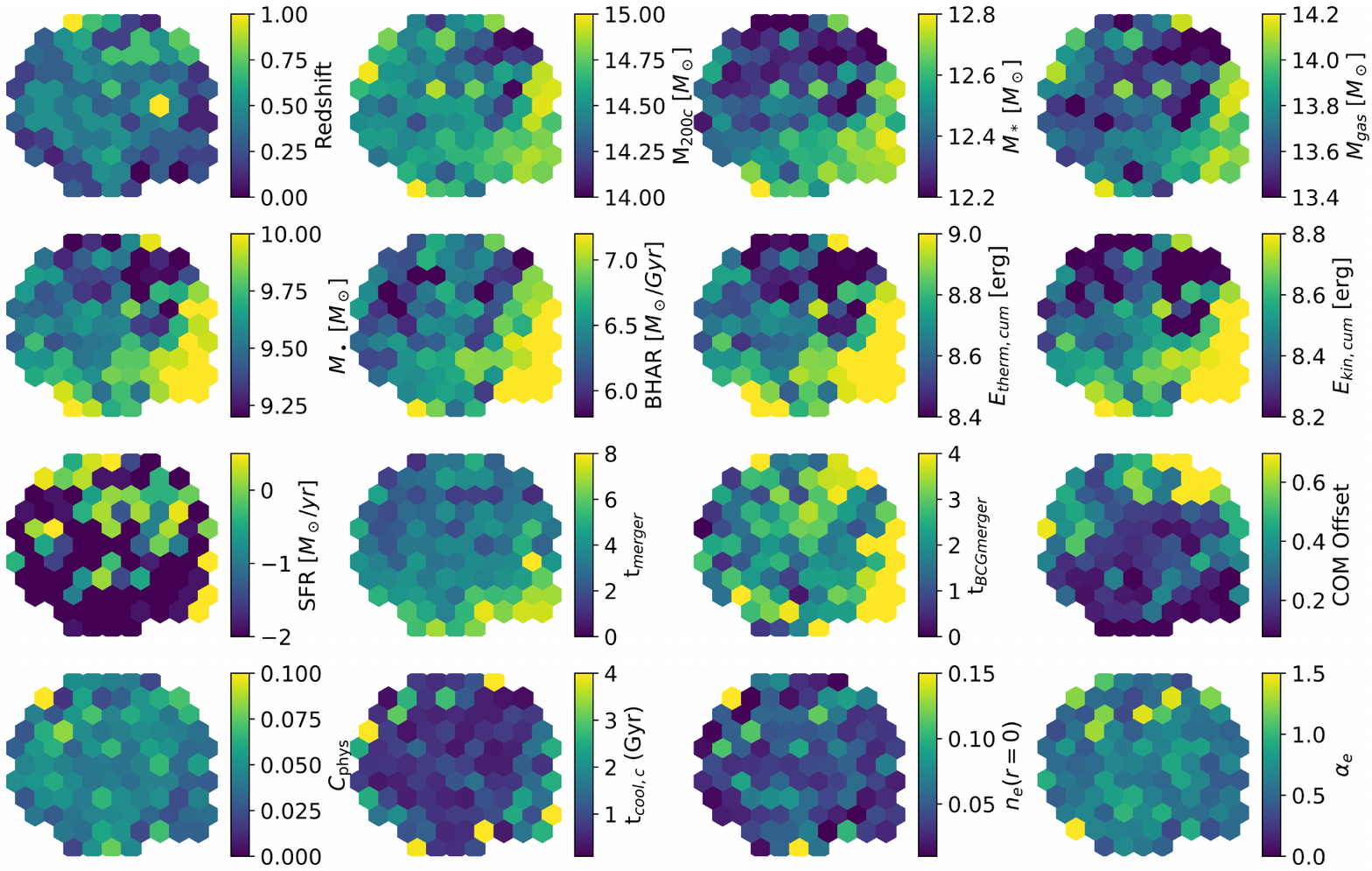}
    \caption{\textbf{Demonstration of the meaningful relationships between the representation of the clusters maps and their underlying physical properties.} We show hexbin histograms of TNG-Cluster test objects in the 2D UMAP parameter space each color-coded by the median cluster property in each bin, for the properties listed and defined in Table \ref{tab:props}. The UMAP coordinates, and therefore the representation upon which it is based, are clearly related to physical properties of the clusters. For example, the visibly merging clusters on the top of the UMAP from in Fig.~\ref{fig:sorted-images} indeed have had a more recent merger, as indicated by the lower $t_{\rm merger}$ and $t_{\rm BCGmerger}$; on the other hand, the isolated, cuspy clusters to the bottom right show the longest times since last merger. Merging clusters are preferentially at higher redshift. All the mass related properties follow a gradient similar to that of the halo mass. The BCGs of the clusters at the bottom of the UMAP are preferentially quenched. There are no visible trends in the cool-core indicators, whereas both the instantaneous and the cumulative characterizations of the SMBHs activity exhibit clear trends across the UMAPs.}
    \label{fig:umap}
\end{figure*}

\section{Results}
\label{sec:results}

Thanks to TNG-Cluster, we can demonstrate that contrastive learning is capable of sorting (in a self-supervised fashion) thousands of ICM X-ray images: we show this below, together with how galaxy cluster properties distribute in the ensuing representation space and with promising downstream applications.

\subsection{Self-supervised sorting of ICM X-ray images}
Fig.~\ref{fig:sorted-images} shows the results of the NNCLR sorting of X-ray cluster-wide images from the test sample from TNG-Cluster simulation suite. 

We divide the UMAP of the learned representation into a 24 $\times$ 24 grid; we then show the image of one randomly-chosen galaxy cluster in each grid cell. As noted in \S \ref{sec:tng}, each image has a side of length $4R_{\rm 200c}$ and constitutes a projection of depth $2R_{\rm 200c}$ along the line of sight.
As noted in \S \ref{sec:nnclr}, the images are all self-normalised, such that the 99-th percentile brightest pixel in the original image becomes 1, and all values 4 orders of magnitude fainter map to 0. This means that if an image appears brighter overall, it does not imply a brighter galaxy cluster, but rather a flatter gas profile. 

In Fig.~\ref{fig:sorted-images} we see a continuum from merging galaxy clusters on the top  to relaxed objects on the bottom, and from radially cuspy profiles on the bottom right to flatter profiles in the bottom left. The number of visually separated merging components decreases as we move perpendicularly away from the top edge of the UMAP, leading to Bullet-type clusters like Abell 2146, Abell 520 and Abell 2744 and then smoothly to single clusters. Mergers with similarly-sized components, on the top left, live close to the ``flat-core'' population on the bottom centre and bottom left, while those with very different component sizes, on the top right, are closer to the cuspy cores. The merging, flat-core and cuspy-core populations are not disjoint, and we tested that this is not a result of the choice of UMAP parameters. 

In general, the distance between pairs of images in this representation space reflects their degree of similarity. We find that the distribution of distances between different projections of the same cluster, and the distribution of distances for random pairs of images, are nearly identical \ref{fig:rep-distance}. This implies that orthogonal projections have very different representations, and, conversely, that some properties, which are intrinsically independent of viewing angle - e.g. time since last major merger, or the thermodynamic properties of the core - may be predicted poorly from projected images.

\subsection{Cluster property trends in the representation space}

\begin{figure*}
    \centering
    \includegraphics[trim=0cm 1cm 0cm 0cm, clip, width=\textwidth]{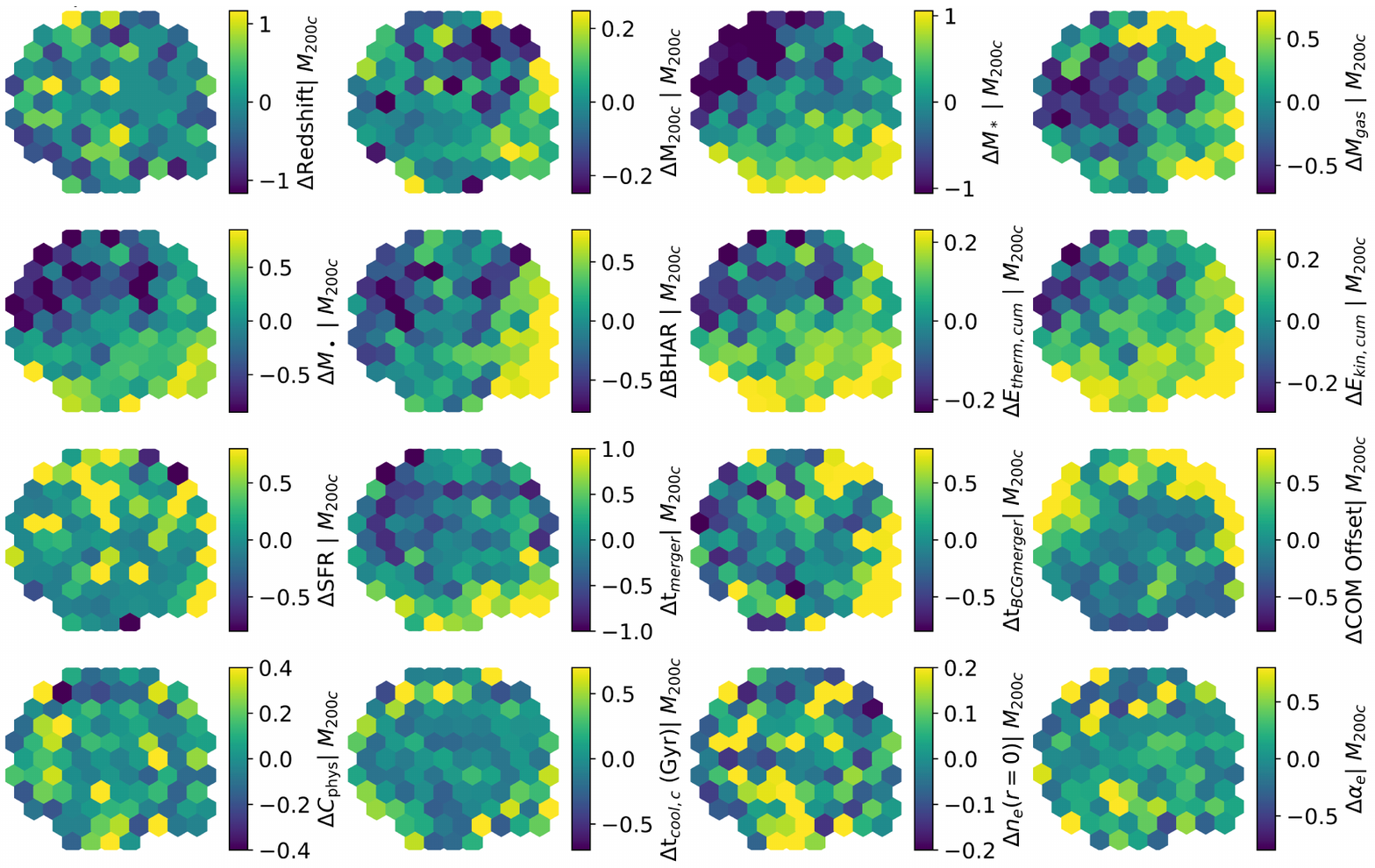}
    \caption{Similar to Fig.~\ref{fig:umap}, but now with scaled properties that are always normalized by the relevant mass or quantity. Since the halo masses grow with cosmic time, this plot still contains some redshift dependence. That said, at fixed halo mass, clusters to the right of the UMAP have higher instantaneous BHAR; those to the bottom and right have higher cumulative energy injections from the SMBHs. Those to the top of the UMAP are more dynamically disturbed. The trends in the cool-core indicators do not track one another.}
    \label{fig:umap-mnorm}
\end{figure*} 

Fig.~\ref{fig:umap} shows median values of various galaxy cluster properties in the UMAP of Fig.~\ref{fig:sorted-images}. And building upon this, in Fig.~\ref{fig:umap-mnorm} we show the same properties but scaled and normalized so that the dependence on total cluster mass is removed: namely, for any given property, we measure the deviation for each cluster from the mean property in a narrow bin of total cluster mass ($\Delta M_{\rm 200c}$ = 0.1 dex), normalized by the standard deviation of the property, and we color-code the UMAP hexbins by the median of such quantity across the maps that fall in each hexbin. 

The merging galaxy clusters on the top left of the UMAP are preferentially $z > 0.5$. The halos of the clusters that visually appear to be merging have indeed undergone a merger with a halo of comparable mass (total mass ratio  $> 0.2$) more recently than those that appear visually relaxed, i.e. they have lower values for $t_{\rm merger}$. The correlation with time since recent merger of the BCG ($t_{\rm BCGmerger}$) is similar, but more noisy. This is expected, since it can take many dynamical times (billions of years) for BCGs to merge after the virial radii of their host clusters first overlap.  

Merging clusters with high mass ratios (i.e. similar component masses, to the top left of the UMAP) have preferentially lower stellar and SMBH masses at fixed halo mass than the more relaxed clusters. The cumulative energy injection from the SMBHs follows a trend similar (albeit not identical) as SMBH mass. The instantaneous SMBH accretion rate (BHAR) is particularly enhanced in a small region to the bottom right of the UMAP, where the star formation rate is also higher than average. This region also hosts the clusters with the longest time since last major merger, $> 7$~Gyr. Visually, these clusters have the most centrally-concentrated i.e. cuspy surface brightness profiles.

\begin{figure*}
    \centering
    \includegraphics[trim=0cm 0.7cm 0cm 0cm, clip, width=\textwidth]{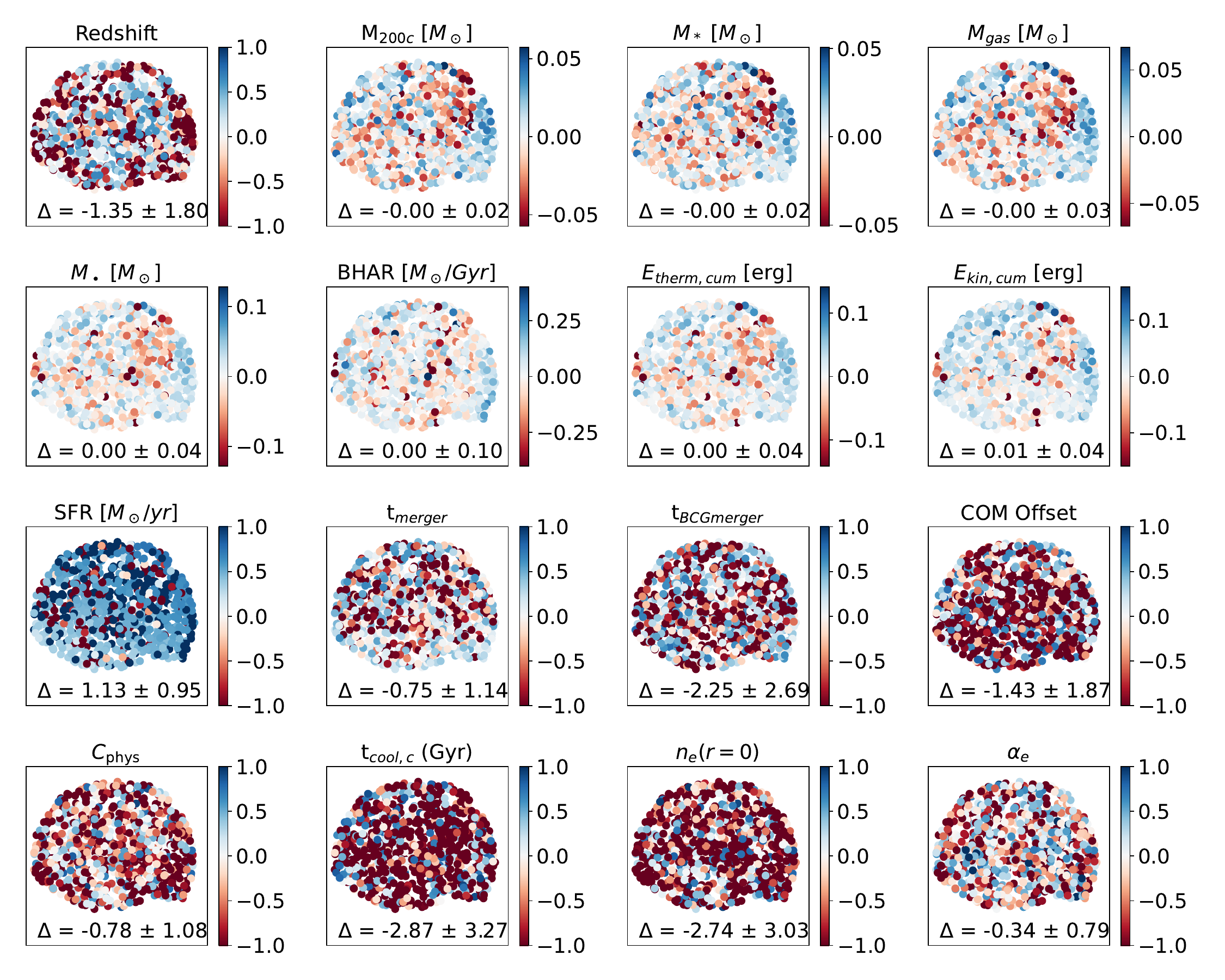}
    \caption{\textbf{Is it possible to infer physical properties of the galaxy clusters by simply accounting for their positioning in representation space?} Here we show the fractional error in the cluster properties predicted by fitting a simple plane to the UMAPs of Fig.~\ref{fig:umap}, i.e. Eq. \ref{eq:fitplane}. The text in each panel shows the mean and standard deviation of this prediction error. Halo, stellar, gas, and SMBH mass, as well as instantaneous BHAR and cumulative SMBH energy injections are all predicted at the few percent level for all but 2-$\sigma$ outliers, i.e. for 96 percent of the images. The remaining properties are not well described by a plane, either because the trends are more complex or because there is no clear global trend. The colorbars are capped at $\pm 100$ percent for the poorly-predicted properties.  }
    \label{fig:predict}
\end{figure*}

% \begin{figure*}
%     \centering
%     \includegraphics[width=\textwidth]{predict_from_umap_repdim8_m200norm.png}
%     \caption{Same as Fig.~\ref{fig:predict}, but with all quantities normalised by halo mass.}
%     \label{fig:predict-mnorm}
% \end{figure*}

Whether the clusters are merging or relaxed correlates well with the ``relaxedness'' criteria of \citet{Ayromlou2023}. Specifically, the offset between the center of mass and the most bound particle (COM Offset), as a function of virial radius $R_{\rm 200c}$, is larger for the merging images. We also find (but do not show) that the fraction of the halo mass enclosed in the primary ``subhalo'' (also a measure for ``relaxedness'') is lower for the merging systems. All this is understood by the fact that these two properties are almost perfectly anti-correlated and both are highly (anti-)correlated with mergers. 

We do not see any trends in the four indicators of cool coreness from \citet{Lehle2024}, which in turn were chosen from the vast observational literature on the topic. This is despite the fact that, visually, we do see that the cuspy cores (naively expected to be cool-cores) are concentrated to the bottom right of the UMAP, whereas the flatter ones (naively expected to encompass non cool-cores)  are to the left. However, it is worth remembering at this point that the images contain 256 pixels a side for a region of $4R_{\rm 200c}$ around each halo center. A single pixel in each image of Fig.~\ref{fig:sorted-images} thus shows 0.015$R_{\rm 200c}$, while the core-coolness indicators are typically computed at 0.008-0.26 $R_{\rm 200c}$. In other words, these metrics are measured within 0.5-2 pixels of the training images. This reminds us that the training images must have sufficient resolution for the features or properties that we hope to learn about.
 
% \begin{figure*}
%     \centering
%     \includegraphics[width=\textwidth]{correlation_angles.png}
%     \caption{The correlation angle $\theta_{12}$ between pairs of galaxy cluster properties, as defined in Eq \ref{eq:fitangle}. A correlation angle of 0 means that the gradients in Fig.~\ref{fig:umap} point in the same direction, i.e. the two properties increase or decrease together. $\theta=180^\circ$ means that one property increases as the other decreases. We emphasis that this is not the correlation between properties in a cluster catalog; rather, it indicates how similarly different properties evolve in the representation space of galaxy cluster images.}
%     \label{fig:correl}
%% UC: I decided not to show this plot because it is a poor indicator of correlation when the properties don't smoothly vary across the plane, and leads to false conclusions.
% \end{figure*}

\subsection{Predicting cluster properties from the image representation}

If physical cluster properties show such clean trends in the UMAP, could they be predicted from these very low-dimensional image representation space? The simplest fit we could perform, directly in the UMAP space, is:

\begin{equation}
    \label{eq:fitplane}
    p_{\rm fit}(u_1, u_2) = a\times u_1 + b \times u_2 + c,
\end{equation}

where $u_1$ and $u_2$ are the two UMAP coordinates, a, b, c are coefficients defining the plane in this 2D space, and $p_{\rm fit}$ is the physical quantity (e.g., BHAR) we aim to predict for the cluster.

Fig.~\ref{fig:predict} shows that the 2-dimensional fit above predicts all the mass related properties, as well as instantaneous BHAR, at the few percent level. The time since the halo merger is predicted to within an order of magnitude. For the remaining properties, the scatter in the prediction error is similar to the mean error, and the errors are 1-2 orders of magnitude. Simply put, a linear model of the UMAP of this representation cannot predict those properties.  

\begin{figure}
    \includegraphics[width=0.5\textwidth]{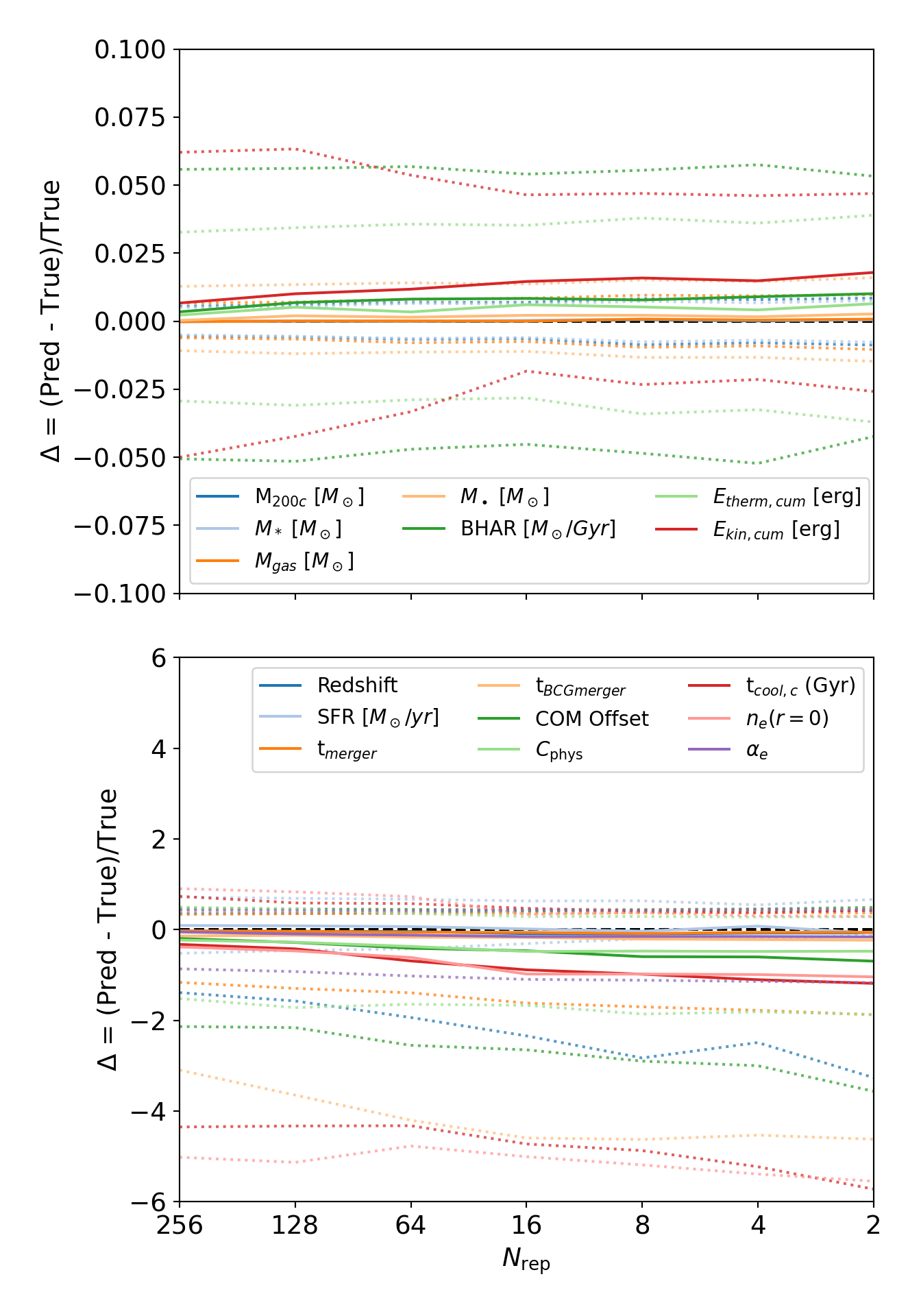}
    \caption{\textbf{Dependence of the prediction results on the dimension of the representation space.} We quantify the fractional error in galaxy cluster properties predicted from the UMAPs using Eq \ref{eq:fitplane}, as a function of the inherent dimensionality of the representation. The default choice throughout the paper is 8 and here we provide tests assuming different choices. The top panel shows properties that the 2D UMAP can predict with a $1-\sigma$ accuracy better than 10 percent. Solid lines mark the median of the prediction errors, while the 16th and 84th percentiles across the cluster sample are marked with dotted lines in the same colour. These properties all scale with halo mass, and could be described well with just one parameter. The bottom panel shows the remaining properties, which are more poorly predicted. These could be better described with higher-dimension representations, but are predicted at best with one order of magnitude accuracy even for N = 256.}
    \label{fig:predict-ndim}
\end{figure}

Fig.~\ref{fig:predict-ndim} shows the median and standard deviation of the error in the predictions using the UMAPs from models, i.e. representation spaces, of increasing dimensionality. The top row focuses on cluster properties for which the $1-\sigma$ range of the prediction error is less than 10 percent, while the bottom panel shows the rest. Almost all the properties benefit slightly from larger representations, since the median error drifts away from 0 for lower N. For the kinetic energy injection, however, the $1-\sigma$ range of error increases marginally (from 5 to 7 percent) for N $>$ 16.

All this is done by looking for quantitative trends in the 2-dimensional UMAP. We checked that a linear model such as that of Eq.~\ref{eq:fitplane} does not perform better when directly fitted to the full N-dimensional representations, for N = (2, 256). In fact, Fig.~\ref{fig:rep-prop-correl} shows that these properties do not correlate strongly with any of the representation parameters. This is also the reason why we chose to opt for a 2D fit to the UMAPs of higher-dimensional representations (Fig.~\ref{fig:predict-ndim}) rather than the full N-dimensional representation: we do not know the geometry of this space, and so it would not be possible to identify a priori a multi-dimensional model.

Indeed, it is crucial to note that the ResNet has been trained in a self-supervised way, without knowledge of the properties, and therefore the learned representations do not know about the physical properties. The need to be passed through the ResNet, convolved with kernels of the trained weights and biases to reproduce the images that the ResNet was trained on. The ResNet never saw these physical properties. 

We intend to infer physical cluster properties from the clusters' maps representation space via neural networks in subsequent papers.

\begin{figure*}
    \centering
    \includegraphics[width=\textwidth]{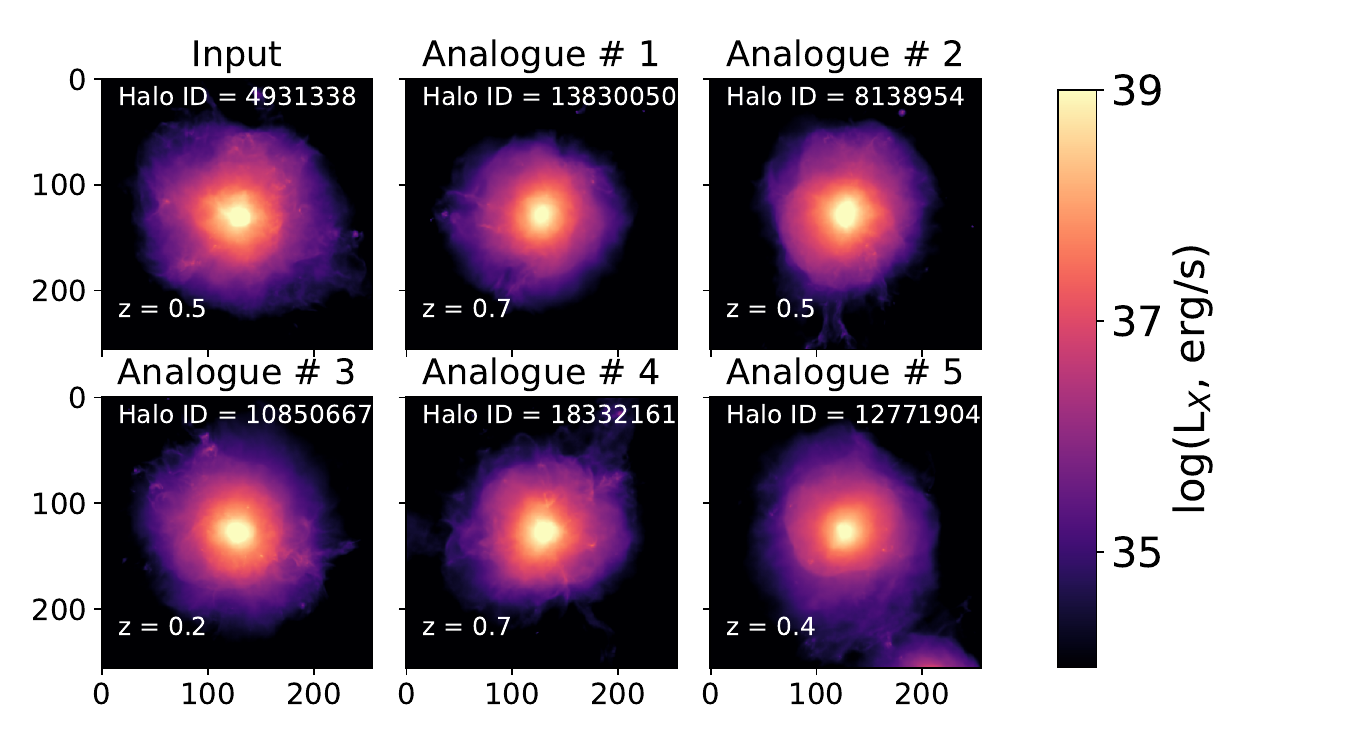}
    \includegraphics[width=\textwidth]{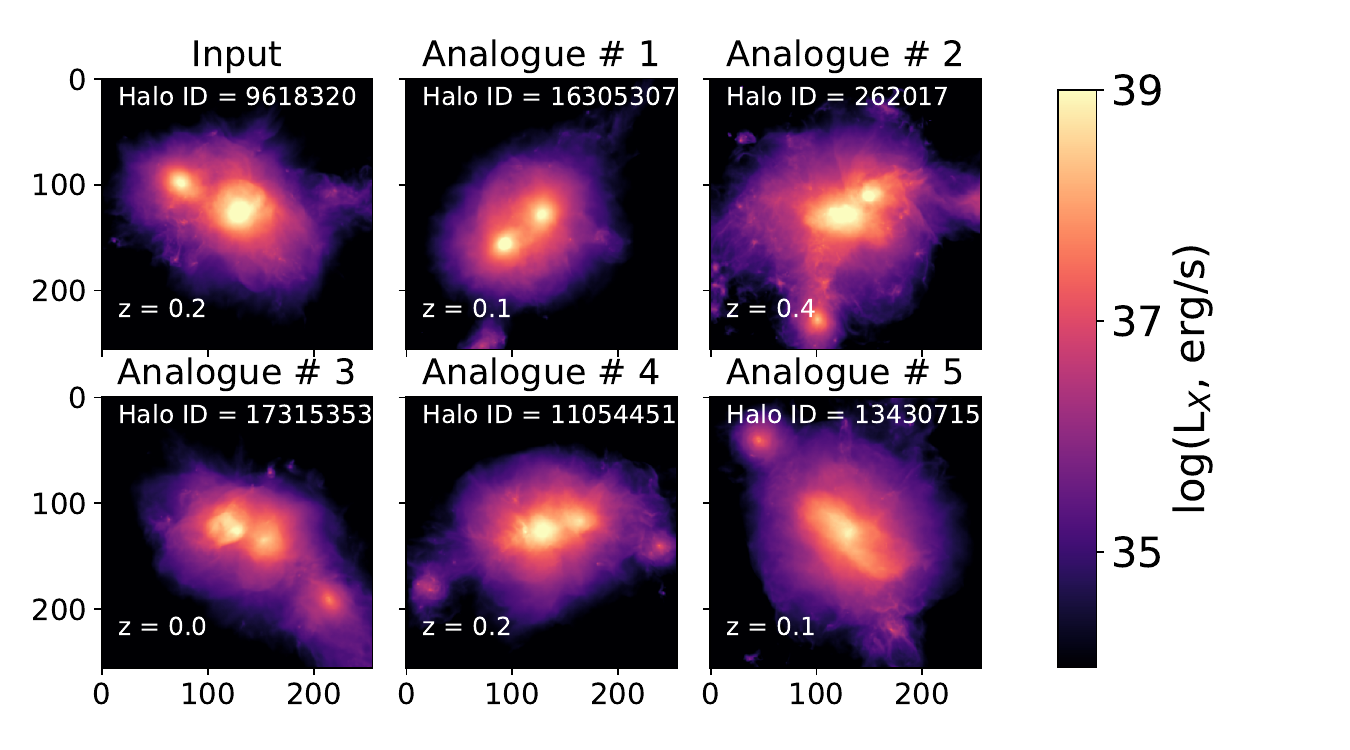}

    \caption{\textbf{Finding analogs with contrastive learning based on X-ray maps.} We show the five nearest neighbours of two example target images, a relaxed cluster (top rows) and a merging cluster (bottom rows), whereby the target image is the top left map of each set. For the case of the merging system, the analogues are all low-redshift clusters with two major merging components, and a third (or sometimes more) smaller components infalling at the outskirts. This showcases the power of the NNCLR methodology applied to astronomy images for the purposes of finding counterparts. %On the \textit{right} we quantify how the true physical properties of the example target cluster from the left (shown in red) compare to a) the 1-$\sigma$ range of each property across all the 352 TNG-Cluster halos (sampled, as in the rest of this study, at 10 snapshots for 0 $\leq$ z $\leq$1; n grey) and b) the corresponding range for the 150 nearest neighbours in the representation space (in blue).
    }
    \label{fig:find-analogs}
\end{figure*}

%\begin{figure*}
%    \centering
%    % 
%    \includegraphics[width=\textwidth]{relaxed_halo_rep_pdf.pdf}
%    \caption{Same as Fig.~\ref{fig:find-analogs}, but for a relaxed halo.}
%    \label{fig:find-relaxed-analogs}
%\end{figure*}

% \begin{figure}
%     \centering
    
%     \caption{}
%     \label{fig:merging-analog-pdf}
% \end{figure}

\begin{figure*}
    \centering
    \includegraphics[width=\textwidth]{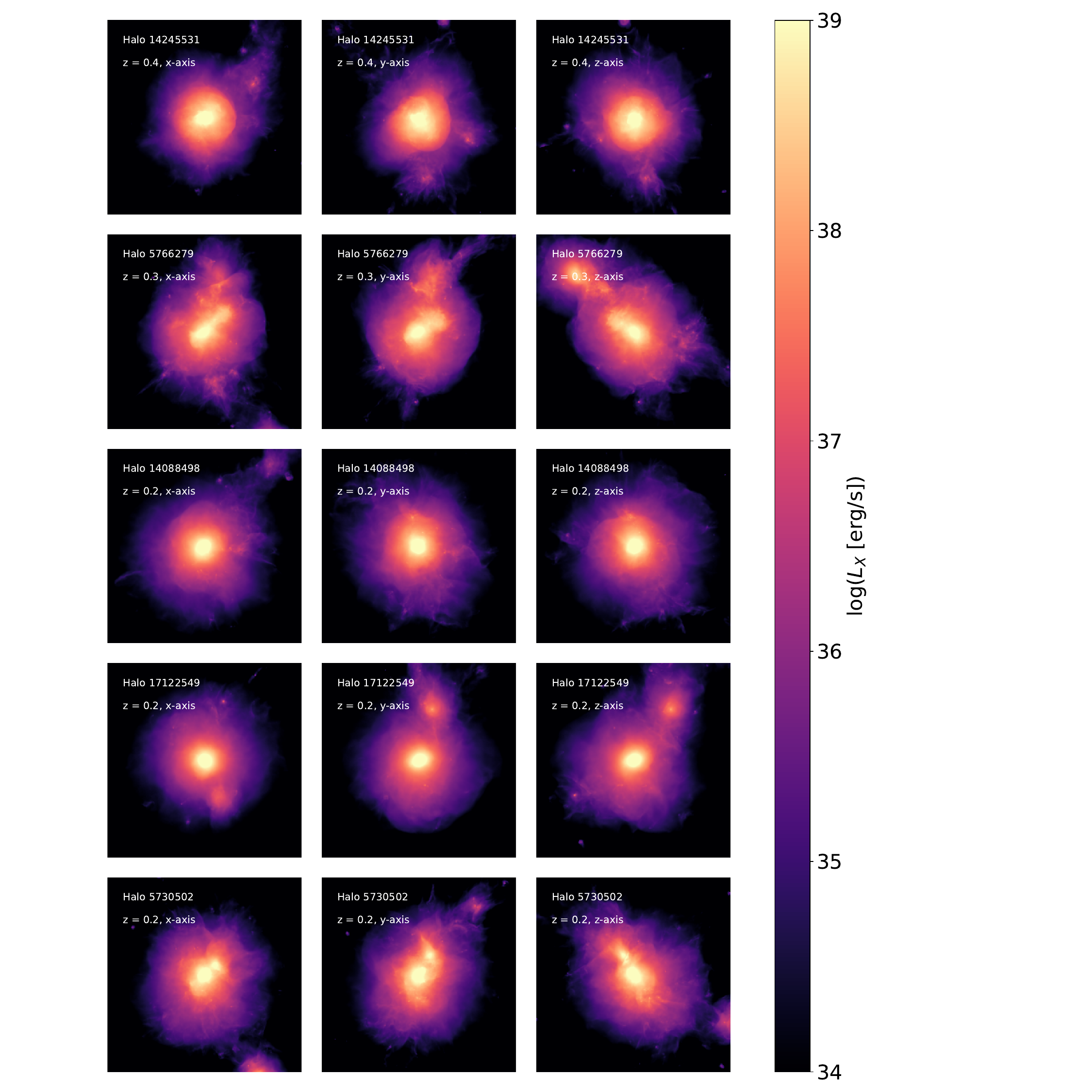}
    \caption{\textbf{Searching for analogs by inspecting the physical properties output from cosmological simulations of clusters}. Here we show images of clusters from the TNG-Cluster simulation at times where the merger tree indicates a merger of mass ratio $> 1:5$ and impact parameter $< 500$ kpc. Many of these appear very relaxed, without multiple visible components, just because of projection effects. Even though such a search would return actually merging system, it may not be the most accurate way to look for analogues of e.g. an observed merging cluster.}
    %\ap{TODO: reduce white margins throughout}}
    \label{fig:find-mergers-tree}
\end{figure*}

\subsection{Finding analogues of observed galaxy clusters}
Another way of constraining the physical properties of galaxy clusters is by looking at the distribution of the properties of its neighbours in the representation space. In fact, this is a corollary of one of the major applications of contrastive learning, which is analogue detection. 

Finding analogues of systems, in particular simulated analogues of observed ones, is crucial to understand the histories and physical properties of the latter. This has been especially a challenge (and a focus) in the case of galaxy clusters that are far from equilibrium, as in the case of mergers. On the one hand, and thinking about cluster merging systems, a possible and traditional approach is to explore a vast parameter space of halo masses, angular moment and geometries with a suite of idealised simulations \citep[e.g.][]{ZuHone2011, Maurogordato2011, Bruggen2012, Machado2013, Chadayammuri2022}. While insightful, this method is limited by the vastness of the parameter space, which grows exponentially with the number of merging components, and does not include priors on any of the quantities, which are naturally reproduced in a cosmological volume. On the other hand, given a cosmological simulation, one can search for analogues of an observed system by either using known physical properties or by inspecting visually observed-like maps. The latter approach is clearly time consuming and, in the case of galaxies, it has been achieved more easily thanks to citizen-science initiatives \citep[see e.g. the identification of jellyfish galaxies from the TNG simulations,][]{Zinger2024}. For galaxy clusters, the visual inspection of the outcome of cosmological simulations for the purpose of analogue finding has been limited so far by the lack of large numbers of galaxy clusters simulated in a cosmological setting \citep[but see][and their figures 3 and 4]{Nelson2024}.
%Until the advent of TNG-Cluster, this approach was limited simply because of the lack of galaxy clusters simulated in a cosmological setting. 

In the following we showcase the power of contrastive learning to identify analogues of given clusters, paving the way towards application with observed clusters.

Fig.~\ref{fig:find-analogs} shows the 5 nearest neighbours in the representation space of two TNG-Cluster examples, a relaxed cluster (top) and a merging one (bottom). We also compare, but do not show, the physical properties of the targeted cluster, the distribution of the properties of the 150 nearest analogues and of the entire TNG-Cluster population. Not only contrastive learning is an excellent tool to find analogues at the map level, as visually conveyed by the maps, but we also find that, for most of the properties explored in this paper, the analogues detected as neighbours in the representation space have physical properties that are more similar to each other than to the entire population.

The application of analogue finding via contrastive learning seems particularly promising for e.g. merging clusters, and overall a more direct tool than more traditional ones. As a reference point and alternative method we show in Fig.~\ref{fig:find-mergers-tree} the result of a search of simulated merging clusters without relying on ML. Namely,  guided by idealised simulations or observation-based arguments, we could apply certain cuts on the mass ratios and impact parameters of mergers identified by particle-based merger trees like SubLink \citep{RodriguezGomez2015} and LHaloTree \citep{Springel2005}. Fig.~\ref{fig:find-mergers-tree} shows images from TNG-Cluster where the mass ratio of the merging halos is $>$ 0.2 and the impact parameter is $<$ 500 kpc, as example choices. Note that the same cluster may appear different from different viewing directions, since the intrinsic merger properties are the same. Some of the selected clusters - e.g. TNG-Cluster with ID 3818655 - do not visibly appear to be merging along any of the chosen projections, while the others are all in very different stages of merging. So, whereas such a search would return actually merging systems, it may not be the most accurate way to look for analogues of e.g. an observed merging cluster. Depending on the scientific application, a combination of approaches might be warranted.

%For finer constraints on a particular system, one could run a suite of simulations exploring the parameter space initially constrained in this manner.  

%%%%%%%%%%%%%%%%%%%%%%%%%%%%%%%%%%%%%%%%%%%%%%%%%%%%%%%%%%%%%%%%

\section{Discussion}
\label{sec:discuss}

\begin{figure*}
    \centering
    \includegraphics[height=1.2\textwidth]{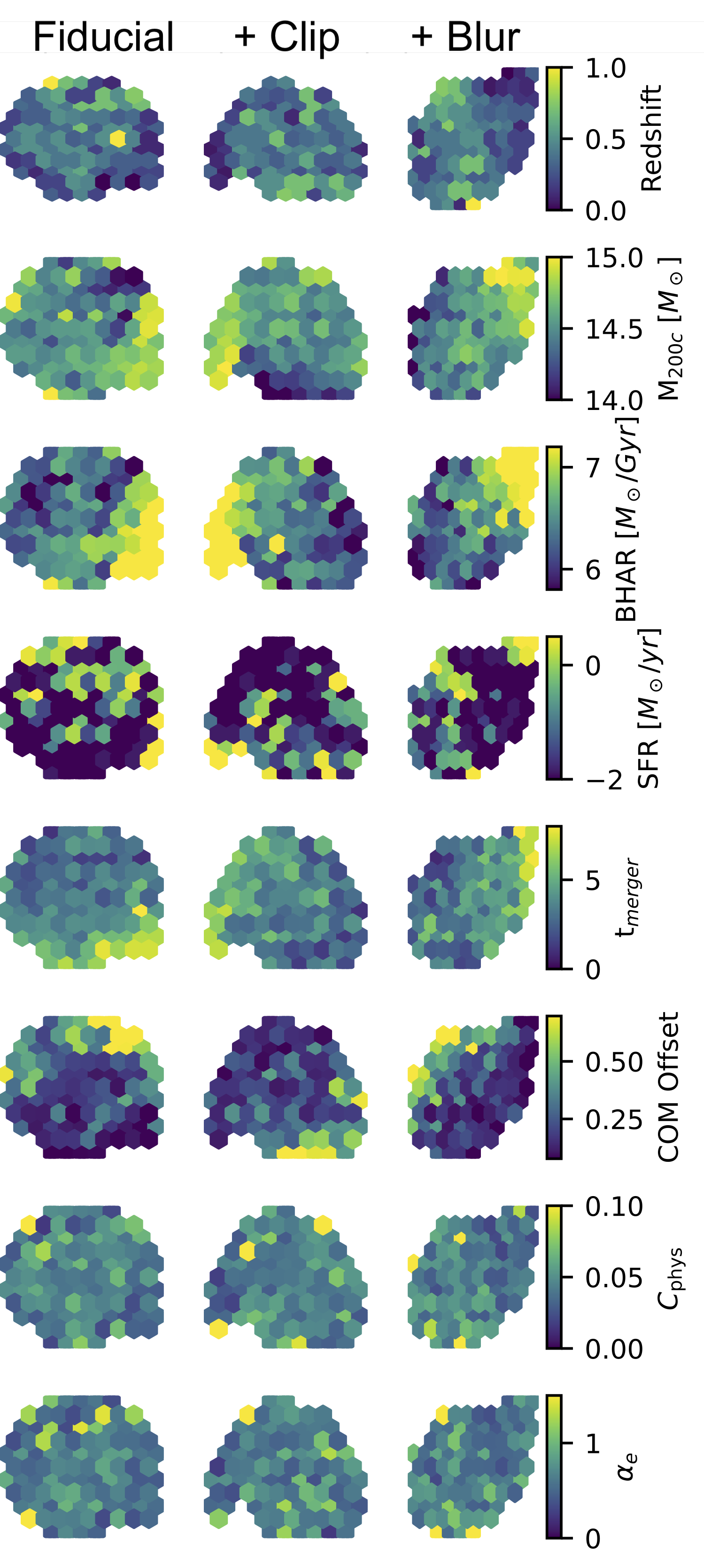}
    \caption{\textbf{Robustness of the representations of cluster X-ray maps against depth (i.e. exposure time) and spatial resolution.} We compare the UMAPs of Fig.~\ref{fig:umap} in our fiducial NNCLR model to alternative models that include either an augmentation for surface brightness clipping (center) or Gaussian blurring (right). The clipping augmentation indicates that even if the faintest regions of a cluster are not observed, the representation is similar. The Gaussian blurring augmentation ensures that the same cluster viewed at different resolutions has similar representations. The exact values of the UMAP coordinates, and the shape of the UMAPs, are different, but the relationships or trends of different properties are preserved across models: the methodology presented in this paper can therefore be robustly applied also to observations with different exposure time or angular resolution.}
    \label{fig:clip-norm}
\end{figure*}

\subsection{What is the minimal effective description of the ICM?}

Fig.~\ref{fig:predict-ndim} suggests that properties that scale with mass can be well described with just a 2D representation. On the other hand, more complex properties indicating star formation, merger activity and core thermodynamics benefit from larger representations, although predictions are not better than order of magnitude for even N=256. However, the fiducial 8-dimensional representation is sufficient at finding visually similar analogues even for merging galaxy clusters, a population which is particularly difficult to characterise from observations. 

The study of representation spaces is beyond the scope of this work, but it is a promising extension of it. A particularly interesting avenue is designing the ResNet to have an Information-Ordered Bottleneck \citep{Ho2023}, which will naturally rank-order the representation dimensions by importance and simplify the process of finding the optimum representation size.

\subsection{Can we recover informative representations with realistic observations of the ICM?}

This study is based on X-ray emission from all the gas associated with a galaxy cluster in the simulation. The images have spatial resolution of a few kiloparsecs ($4\times R_{\rm 200c}/256$, the latter being the pixel per side) and a dynamical range of four orders of magnitude; on the other hand, realistic observations would have worse spatial resolution, PSF convolution effects, and a lower dynamical range, scaling with instrument sensitivity and exposure time. 

In fact, both these limitations can simply be implemented as augmentations in the training pipeline. We can teach the model to learn a representation of the ICM that does not care if a telescope observed it for 10~ks or 1~Ms, because, if the images are normalised to the brightest pixel, this is equivalent to setting all the pixel values below a certain threshold to 0. Note that this is different from training a variety of models, each using images clipped at a different minimum brightness. Instead, at each instance of computing contrastive loss, the two images of the pair will have a randomly applied clipping augmentation. Similarly, a blurring augmentation enforces the fact that a galaxy cluster is the same whether observed by Chandra (with a PSF of 1") or eROSITA (25"), because we can convolve the image with Gaussians of different widths and treat differently blurred images as a positive pair.

Fig.~\ref{fig:clip-norm} shows the effect of adding a clipping (central column) and blurring (right column) augmentation to the fiducial model described here. The UMAPs are oriented differently because of the random nature of the transform; we should focus instead on the gradients in different properties across the UMAP and compare them to each other. As in the fiducial model, we see that the regions of highest halo mass show the highest BHAR, and the most extreme values of BHAR correspond to high SFR. The largest COM Offset is always in areas with the lowest time since a major merger, $t_{\rm merger}$. There are no trends in the cool-core indicators for any of the models; as already noted, this is due to the resolution of the images used in this study.

For an end-to-end and robust application to observations, we would recommend training a model with faithfully-mocked images that account for instrument effective area, chip gaps, and background noise. Such studies should be tailored to the observational sample at hand, and are beyond the scope of this work. 

% \subsection{Interpreting the representation space}
% A common concern about neural networks is the idea that they are ``black boxes", difficult if not impossible to interpret. This is far from true. On the one hand, we can study the topology of the latent (a.k.a. representation) space, and of how the data fill that space \citep{Wu2017, LopezGonzalez2024}. Second, we can modify the training such that the coordinates in the representation space are sorted by importance \citep{Finn2014, Ho2023}. By training a U-net or autoencoder style architecture rather than only an encoder, we can identify the role of each latent representation component by feeding in a vector to the decoder that has only one non-zero element, i.e. feeding in only one representation component at a time, and analysing the output images. 

%%%%%%%%%%%%%%%%%%%%%%%%%%%%%%%%%%%%%%%%%%%%%%%%%%%%%%%%%%%%%%%%

\section{Summary and Conclusions}
\label{sec:concl}

In this ERGO-ML paper (Extracting Reality from Galaxy Observables with Machine Learning), we use nearest neighbour contrastive learning (NNCLR) to find a low-dimensional representation of images of the X-ray emitting gas from galaxy clusters. In particular, we develop and showcase a new method to study populations of galaxy clusters by using the output of the TNG-Cluster suite of cosmological magnetohydrodynamical zoom-in simulations. The training sample consists of images of the intrinsic X-ray emission from all the gas cells within a cube of size $4\times4\times2 R_{\rm 200c}$, with the shortest axis being the line of sight, and with pixel sizes of 0.015 $R_{\rm 200c}$ i.e. $ \sim$ 10-20 kpc. Such a training sample consists of nearly 6,200 images, and the test and validation samples consist of 772 images each. We find that:

\begin{itemize}
    \item Contrastive learning produces a semantically-meaningful representation of the X-ray images of galaxy clusters that would agree with a human description of similarity. When the ensuing representation space is visualised as a 2D UMAP as in Fig \ref{fig:sorted-images}, we see smooth trends from e.g. complex cluster mergers with well-separated components to Bullet-like mergers close to pericenter passage and from flat, non-cool core clusters to isolated clusters with peaked, cool cores.
    \item We find, as seen in Fig \ref{fig:umap} that galaxy cluster properties such as total mass, SMBH, star formation activity, and dynamical state vary smoothly within this representation space. 
    \item These trends are also recovered with surface brightness clipping and Gaussian blurring augmentations, as shown in Fig \ref{fig:clip-norm} making this new analysis method applicable to observational samples with non-uniform exposure times and even across instruments. 
    \item Mass-related properties are described to better than 5 percent accuracy, with no bias, as a linear function of just the 2 UMAP coordinates (Fig \ref{fig:predict}). Instantaneous black hole accretion rate and dynamical state are also predicted with very low bias but with a poorer accuracy of 10 and 17 percent, respectively.
    \item The remaining cluster properties are harder to predict with our deliberately-simple fit of the UMAP of the representations space. By using the full representation space rather than the 2D UMAP, we find in Fig \ref{fig:predict-ndim} that the prediction accuracy improves with higher dimensionality, flattening around N=128. Even so, the cool-core indicators and time since BCG merger cannot be predicted to better than an order of magnitude. These properties are characterised on the scales of 0.01-2 pixels, and could possibly be recovered from a model trained on zoomed-in regions of the clusters core. 
    \item Distances between maps in the representation space can be used to find visually-analogous images, whether they are relaxed or highly disturbed \ref{fig:find-analogs}. This will be particularly powerful when applied to real observations of galaxy clusters, which in turn requires training the contrastive learning model on realistic mocks. We will present this work in a follow-up paper. 
\end{itemize}

Overall, we find contrastive learning to be a very powerful tool in the study of the ICM. The ResNet architecture is capable of characterising the complex extended features in the diffuse ICM, whereas the augmentations make this representation invariant to physically-irrelevant transformations such as rotation, translation, zooming, cropping, addition of Gaussian noise, blurring, and even varying dynamical range. This visual representation is correlated to the mass, SMBH activity, and merger history of the galaxy clusters. It allows us to characterise the rich galaxy cluster population in a smooth fashion, rather than reducing images to scalars and then applying discrete cuts on them. This is particularly useful and exciting given the very large cluster catalogs being produced with e.g. the eROSITA telescope \citep{Liu2022, Bulbul2024}.

\section*{Data Availability}

The IllustrisTNG simulations themselves are publicly available and accessible at \url{www.tng-project.org/data} \citep{Nelson2019}, where the TNG-Cluster simulation will also be made public in the near future. The NNCLR implementation and post-processing code can be accessed at \url{https://github.com/milchada/icm-clr/}. Data directly related to this publication is available on request from the corresponding authors. This work has benefited from the \texttt{scida} analysis library \citep{Byrohl2024}.

\section*{Acknowledgements}

We thank Shy Genel, Yuan Li and Lawrence Rudnick for useful discussions. UC and AP acknowledge funding from the European Union (ERC, COSMIC-KEY, 101087822, PI: Pillepich). MA and DN acknowledge funding from the Deutsche Forschungsgemeinschaft (DFG) through an Emmy Noether Research Group (grant number NE 2441/1-1). KL acknowledges funding from the Hector Fellow Academy through a Research Career Development Award. This work is supported by the Deutsche Forschungsgemeinschaft (DFG, German Research Foundation) under Germany’s Excellence Strategy EXC 2181/1-390900948 (the Heidelberg STRUCTURES Excellence Cluster). The TNG-Cluster simulations have been carried out with compute time under the TNG-Cluster project on the HoreKa supercomputer, funded by the Ministry of Science, Research and the Arts Baden-Württemberg and by the Federal Ministry of Education and Research. In addition, simulations were run on the bwForCluster Helix supercomputer, supported by the state of Baden-Württemberg through bwHPC and the German Research Foundation (DFG) through grant INST 35/1597-1 FUGG; the Vera cluster of the Max Planck Institute for Astronomy (MPIA), as well as the Cobra and Raven clusters, all three operated by the Max Planck Computational Data Facility (MPCDF); and the BinAC cluster, supported by the High Performance and Cloud Computing Group at the Zentrum f\"ur Datenverarbeitung of the University of T\"ubingen, the state of Baden-Württemberg through bwHPC and the German Research Foundation (DFG) through grant no INST 37/935-1 FUGG. 

\bibliographystyle{mnras}
\bibliography{reference}

%%%%%%%%%%%%%%%%%%%%%%%%%%%%%%%%%%%%%%%%%%%%%%%%%%%%%%%%%%%%%%%%

\appendix

\section{On using the same galaxy cluster at different snapshots and viewing directions in the training}

\begin{figure}
    \includegraphics[width=0.48\textwidth]{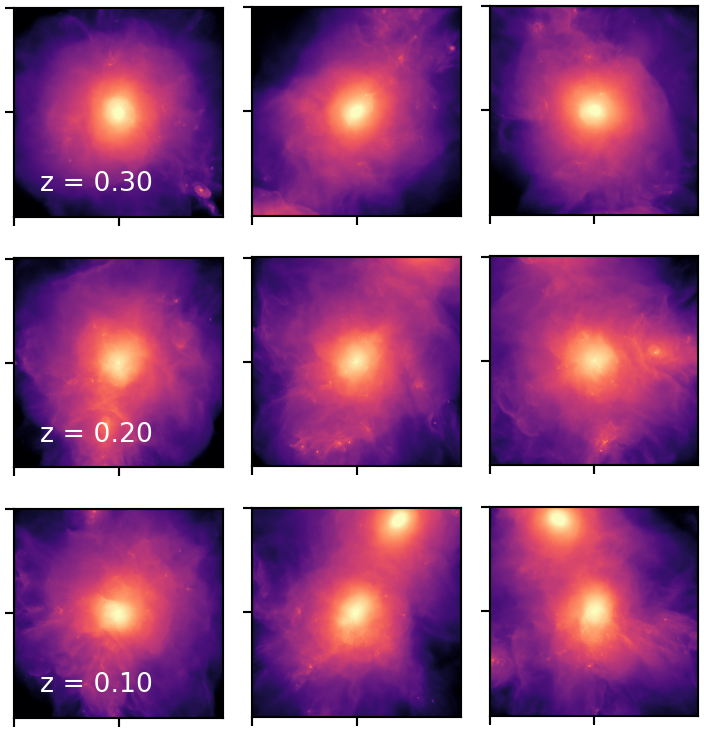}
    \caption{Intrinsic X-ray emission from one galaxy cluster (Halo ID 15341007) at three consecutive full snapshots, projected along the $\hat{x}$ (left), $\hat{y}$ (center) and $\hat{z}$ (right) axes of the simulation box. The nine images are visually very different from each other, and act as independent samples in the training. }
    \label{fig:example-projection}
\end{figure}

\begin{figure}
   \centering
   \includegraphics[width=0.5\textwidth]{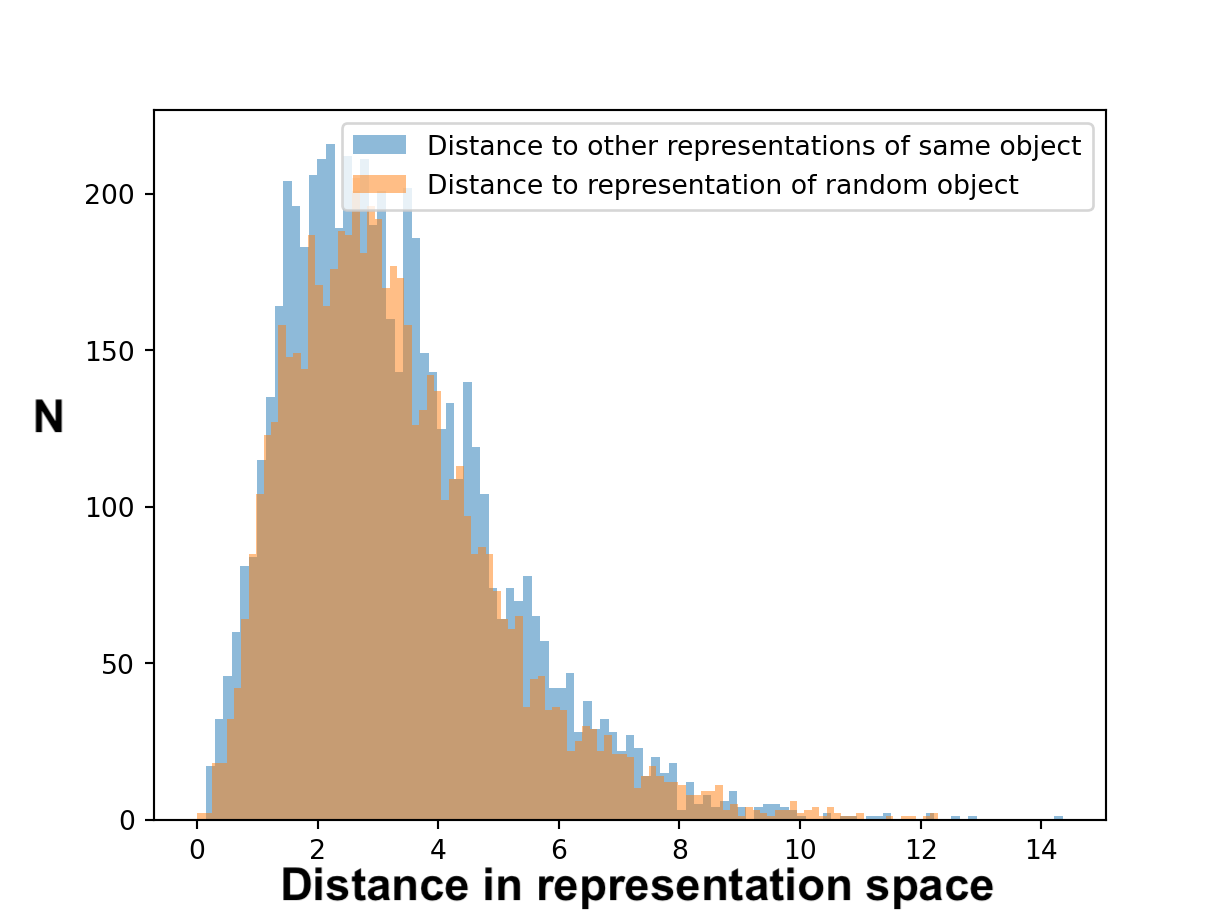}
   \caption{Distance in the representation space between different projections of the same cluster (blue) vs between two random images (orange). The two distributions are almost identical, validating the use of orthogonal projections of one cluster as independent training samples. }
   \label{fig:rep-distance}
\end{figure}

A priori, using duplicates in the training sample due to multiple projections of the same cluster, as well as the same cluster at multiple snapshots may be problematic. Fig.~\ref{fig:example-projection} shows why, in practice, this is not a concern. Galaxy clusters evolve fairly significantly between the full snapshots of the TNG-Cluster output, and look significantly different from different directions. 

\section{Is the self-supervised representation sensitive to NN archictecture choices?}
The representations learned by the NNCLR are the result of passing the images through a ResNet, which is ultimately a convolutional neural network. There are various architectural choices to be made before training a model -- How many convolutional layers to use? How to trim data between convolutions? What kernel sizes to use at each step, and what stride? Is there enough training data to actually learn the representation we want?

\begin{figure}
    \centering
    \includegraphics[width=0.5\textwidth]{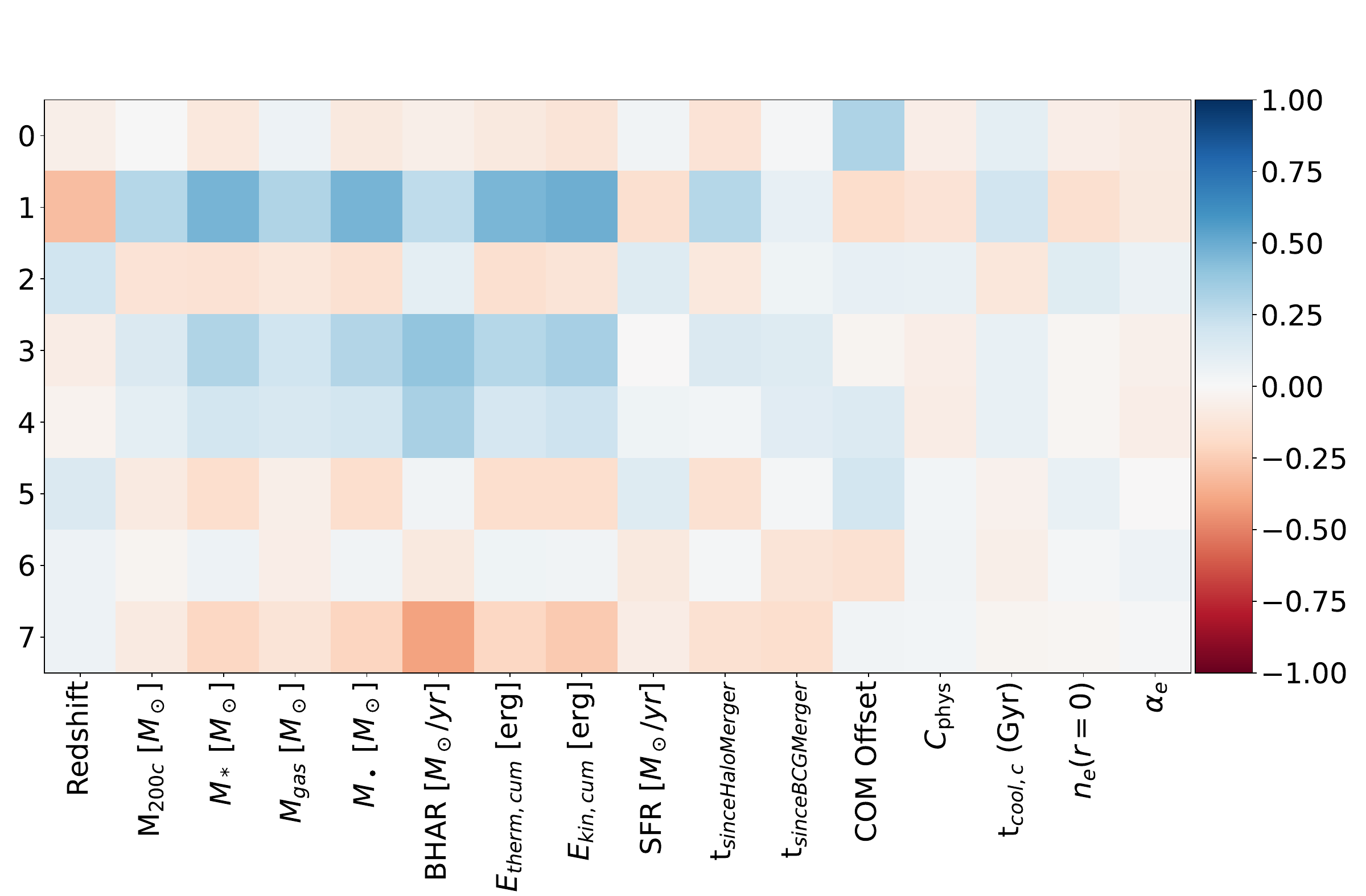}
    \caption{\textbf{Spearman correlation coefficient of the 8 representation dimensions of the maps with mass-scaled physical cluster properties}. No representation dimension correlates precisely with a physical property, but several are moderately correlated to redshift and to SMBH and merger properties.}
    \label{fig:rep-prop-correl}
\end{figure}

\begin{figure*}
    \includegraphics[height=1.3\textwidth]{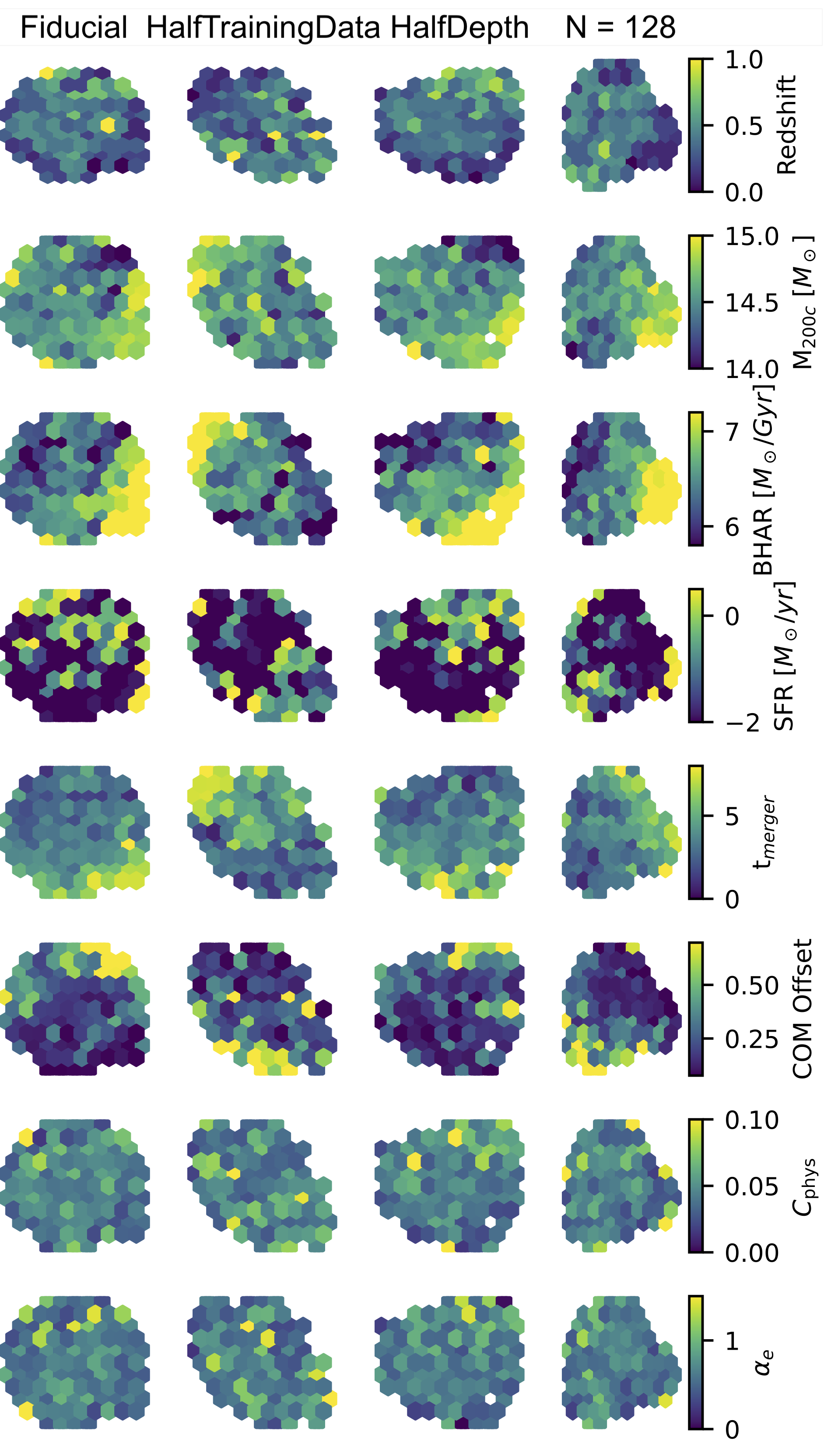}
    \caption{{\bf The effect of varying ResNet model parameters.} The fiducial model adopted throughout this work has 8 dimensions, a training sample of 6,412 images, and a ResNet depth of 16. Qualitatively, all the models show a gradient in physical properties across their UMAPs, with different properties co-evolving (or not) in similar ways. All the models are better at capturing mass and dynamical state than cluster core thermodynamic state.}
    \label{fig:resnet-varied}
\end{figure*}

We use Optuna \citep{Akiba2019} to pick these parameters. Furthermore, we tested models that changed various parameters of the fiducial ResNet architecture, as well as those that used smaller training samples. Fig.~\ref{fig:resnet-varied} shows that all the models reproduced the above trends in galaxy cluster properties in the representation space. 

% \begin{figure*}
%     \includegraphics[width=0.75\textwidth]{images_fiducial_core.png}
%     \caption{Same as Fig.~\ref{fig:sorted-images}, but zoomed in on the central 0.1$R_{\rm 200c}$. Trends are much weaker on this scale, with perhas the only persistent result beng that compact cores without signs of AGN activity remain on the top right. This highlights the imortance of training models on images at the scale appropriate for the question at hand. }
%     \label{fig:image-core}
% \end{figure*}

%\begin{figure*}
%    \includegraphics[width=\textwidth]{rep_phys_corr_repdim=16.png}
%    \caption{Same as Fig.~\ref{fig:rep-prop-correl}, but for representations with 2 (left) and 16 (right) dimensions. Low-dimensional representations primary learn about SMBH mass and activity, followed by stellar and gas mass fractions; additional parameters do not correlate well with physical properties.}
%\end{figure*}

Fig.~\ref{fig:rep-distance} shows the distance between pairs of images in the representation space. The blue histogram shows different projections of the same cluster, while orange is for random pairs of images. We see that the former histogram is only very slightly skewed towards lower distances than the latter. This tells us that orthogonal projections have very different representations, and, conversely, that some properties, which are intrinsically independent of viewing angle - e.g. time since last major merger, or the thermodynamic properties of the core - may be predicted poorly from projected images.

\label{lastpage}
\end{document}